\documentclass[12pt]{article}

\usepackage{lineno,hyperref}
\usepackage[utf8]{inputenc}
\usepackage{morefloats}
\usepackage{amssymb}
\usepackage{bbm}
\usepackage{amsmath}
\usepackage{graphicx}
\usepackage{dsfont}
\usepackage{float}
\usepackage{color}
\usepackage{caption}
\usepackage{subcaption}
\usepackage{color}

\newcommand{\cF}{\mathcal{F}}
\newcommand{\inds}[1]{\mathbbm{1}_{#1}}
\newtheorem{theorem}{Theorem}[section]

\title{A new centered spatio-temporal autologistic regression model. Application to spatio-temporal analysis of esca disease in a vineyard}
\date{2018\\ November}

\author{Anne G\'egout-Petit\\ Universit\'e de Lorraine, CNRS, Inria, IECL, F-54000 Nancy, France, \\ 
anne.gegout-petit@univ-lorraine.fr
\and Lucia Guérin-Dubrana\\, Université de Bordeaux, ISVV, UMR-1065 INRA, France \and Shuxian Li\\
, Université de Bordeaux, ISVV, UMR-1065 INRA, France}
\begin{document}
\maketitle

\begin{abstract}
We propose a new centered autologistic spatio-temporal model for binary data on a lattice. The centering allows the interpretation of  the autoregression coefficients in separating the large scale structure of the model corresponding to an expected mean and the small-scale structure corresponding to the auto-correlation. We discuss the existence of the joint law of the process and show by simulation the interest of this kind of centering. We propose and show the efficiency of the maximum pseudo-likelihood estimator and also a method to choose the best structure of neighborhood. Method is applied to model and fit epidemiological data about Esca disease on a vineyard of the Bordeaux region. \end{abstract}

{\bf Keywords}\\
spatial-temporal modeling; large-scale model structure; binary response; Maximum Pseudo-Likelyhood estimation; Autologistic Model




\section{Introduction}

Since spatial and spatio-temporal data are commonly present in nature, the models on such data had drawn large interests of scientists from various fields such as ecology, epidemiology and image analysis, during past years. Binary data are of particular interest for modelling the occurrence of an event like disease or death so that spatial and spatio-temporal binary data models are very useful to study the evolution of a disease known at each point of a grid if we suspect spread from the past and correlation between neighbors. 

40 years ago, Besag (1974)  in \cite{besag1974spatial} firstly proposed an autologistic model for spatial binary data, assuming a simple dependence on surrounding neighbors. This model was proved a very useful model and then was extended by Gumpertz {\it et al.} (1997) and Heffer and Wu (1998) in order to integrate the regression on the covariates \cite{gumpertz1997autologistic, huffer1998markov}.

More recently, Zhu   {\it et al.} (2008) \cite{zhu2008autologistic}  and Zheng and Zhu (2008) \cite{zheng2008markov} generalized the autologistic regression models to account for covariates, spatial
dependence, and temporal dependence simultaneously for binary data that are measured repeatedly over time on a spatial lattice.

However, the non-centered parametrization of the autologistic regression models present  parameter interpretation difficulties across varying levels of statistical dependence. This problem has been first pointed out by Caragea and Kaiser (2009) \cite{caragea2009autologistic} who proposed a centered parametrization for spatial autologistic regression model to overcome this difficulty.  In the following of \cite{caragea2009autologistic}, Hughes {\it et al.} \cite{hughes2011autologistic} further discussed the estimation and simulations of the autologistic model on lattice under the centered parametrization. Whang and Zheng (2013)  \cite{wang2013analysis}  proposed a centered spatio-temporal autologistic regression  model. They used and compared several estimation methods. The drawback of this spatiotemporal model is that the temporal dependency is not causal and the state of a point at time $t$ is linked with its state at time $t-1$ and $t+1$. This model has good mathematical properties but is not useful for interpretation by the practitioner. A good review about autologistic models for binary data is done in Zhu and Zheng (2009) \cite{zhu2016autologistic}.

In this paper, we propose a new centered spatio-temporal autologistic model which depends only on the past. We will show, by simulations, the advantage of this kind of centering over other  autologistic models. Since the joint distribution of such models are very complex, we propose an inference method based on the  maximization of the pseudo-likelihood. We also proposed a method based on the criterion Pseudo-Likelihood to deal with the choice of the neighbour structure. We apply the model for a better understanding of the spread of esca grapevine disease in a vineyard. We used the foliar symptom data recorded in a vineyard of Bordeaux region from 2004 to 2017.

The paper is organized as follows: in the  following of this section, we present the state of the art about spatial and spatio-temporal autologistic model. After that, in Section 2 we will present our new centered autologistic model and discuss the existence of the joint law of the spatio-temporal process. Then we show several simulations to compare the spatio-temporal autologistic model depending on past under different centered parametrizations in Section 3. Afterward we propose an algorithm of inference by maximization pseudo-likelihood estimation of our model and a method to choose between the possible neighbor's structures and show the efficiency of the method on simulated data in Section 4. Section 5 presents a real data analysis coming from a epidemiological study in vegetal health.  At last Section, we discuss the interest of the methodology and give some perspectives of this study.  

\subsection{Spatial autologistic models}
 Let us use $[Z]$ to denote the distribution of random variable $Z$. Let  $\mathbf{Z}$ be the random field of interest,  $\{Z_i: i=1,...,n\}$; each $Z_i$ being a binary variable indexed by $i$ a point of a spatial lattice $ S = \{1, \ldots, n \}$. The distribution $[Z]$ is given by the conditional laws

$$ [Z_i|Z_j,  j \neq i] \sim Binary(p_i),$$
where for $1 \leq i \leq n$, $p_i=\mathbb{P}(Z_i=1|Z_j,j\neq i)$.

In addition, we have a neighbourhood structure: we assume that corresponding to each location $i$ is a set $N_i$  of other locations considered to be neighbours of $i$. We suppose that this relation is symmetric that is $j \in N_i \Leftrightarrow i \in N_j$. It defines a non-oriented graph whose nodes are the locations $i$'s and there is an edge between $i$ and $j$ if $j \in N_i$. A Markov random field then results from this neighbour structure if we assume that 
\begin{equation} 
[Z_i|Z_j, i \neq j]=[Z_i|Z_j, j \in N_i] \;\;\; \textrm{for all } \; 1 \leq i \leq n
\label{ch6-1} 
\end{equation}

\noindent
The conditional binary probability can be expressed in exponential family form:

\[ \mathbb{P}(Z_i|Z_j, j \in N_i)
=\frac{\exp \left(Z_iA_i(Z_j, j \in N_i)\right)}{1+\exp \left(A_i( Z_j, j \in N_i)\right)}
\]
where $A_i$ is called a natural parameter function. When the $Z_i$'s take values in $\{0,1\}$ and that will be the case in this paper, the use of  the logit function instead of the $A_i$ to model the $p_{it}$ is very common; it is defined by the following equation. 
\[ A_i(Z_j, j \in N_i) = \text{logit}(p_{it})= \log(\frac{p_{it}}{1-p_{it}}) = \log\left(\frac{\mathbb{P}(Z_i=1|Z_j, j\in N_i,}{\mathbb{P}(Z_i=0|Z_j, j\in N_i, }\right).
\]
Besag (1974) \cite{besag1974spatial} showed that the natural parameter functions for this formulation must be of form: 

\begin{equation}
A_i( Z_j, j \in N_i)= \text{logit}(p_{it}) = \alpha_i+\sum_{j\in N_i}\beta_{ij}Z_j.
\label{eq:1}
\end{equation}
with $\alpha_i$ a leading constant. Note that the variables $Z_i$'s are independent if for each $1 \leq i \leq n$, $\mathbb{P}(Z_i=1|Z_j,j\neq i)$ does not depend on the $Z_j$'s that is equivalent to  $\beta_{ij} = 0 \;\;  \forall (i,j) \in S^2$). It means that the parameters $\beta_{ij}$'s measure the dependences inside the lattice. They must satisfy certain restrictions for a joint distribution of the $Z_i$'s to exist \cite{gaetan2008modelisation}.

\noindent
The modelling can take spatial covariates $\mathbf{X}=\{\mathbf{X}_i, i=1,...,n \}$ into account; where for each $1\leq i \leq n$, $\mathbf{X}_i$ is a vector of dimension $p$: $\mathbf{X}_i = (X_{i,1}, \ldots , X_{ip})^T\!\!.$
In this case, the natural parameter function is given in \cite{caragea2009autologistic} by:
\begin{equation}
\text{logit}(p_{i}) =\mathbf{X}_i^T\boldsymbol{\beta}+\sum_{j\in N_i}\beta_{ij}Z_j
\label{eq:2}
\end{equation}Caragea {\it et al.} (2009) in \cite{caragea2009autologistic} discussed the interpretation difficulties for traditional autologistic model given by Equation  (\ref{eq:2}). Indeed, 
let $p_i=\mathbb{P}(Z_i=1|Z_j, j\in N_i, \mathbf{X}_i)$ so that the odds that $Z_i=1$ in model (\ref{eq:2}) is $p_i/(1-p_i)$. Let us denote $c_i= \frac{exp(\mathbf{X}_i^T\boldsymbol{\beta})}{1+exp(\mathbf{X}_i^T\boldsymbol{\beta})}$ being the probability of occurrence under the spatial independence model (when all the $\beta_{ij}$'s equal $0$), the odds that $Z_i=1$ is $c_i/(1-c_i)$. Then the log odds ratio for model (\ref{eq:2}) relative to the independence model is: 

$$ log[\frac{p_i/(1-p_i)}{c_i/(1-c_i)}]=\sum_{j \in N_i} \beta_{ij}Z_j $$

\noindent
In this case,  the odds of $Z_i=1$ in model (\ref{eq:2}) relative to the independence model increases for any nonzero neighbors, and can never decrease. This is not reasonable if most of neighbors are zeros and could bias the realizations towards 1.  \\

\noindent 
To overcome this interpretation difficulties and in a purpose of a better interpretation, a centered spatial autologistic model was proposed by \cite{caragea2009autologistic}. In this model, the value of the $Z_j$'s in the regression are centered by their expected "large-scale" value and the the $p_i$'s are given by
\begin{equation}
\text{logit}(p_{i}) =\mathbf{X}_i^T\boldsymbol{\beta}+\sum_{j\in N_i}\beta_{ij}\left(Z_j- \frac{\exp(\mathbf{X}_j^T\boldsymbol{\beta)}}{1 + \exp(\mathbf{X}_j^T\boldsymbol{\beta})}\right)
\label{eq:2c}
\end{equation}
They pointed out that model (\ref{eq:2c}) is similar to the parametrization customarily used for auto-Gaussian models. This parametrization consider the overall level of a process, possibly adjusted by influence of covariates, to be appropriately modeled as what is called the large-scale model component, while variances, covariances, and other high-order portions of the data structure are accounted for by what is called the small-scale model component (\cite{cressie1993statistics}, p.114). In addition, the parameters were estimated by maximum pseudo-likelihood.

Hughes {\it et al.} (2011) in \cite{hughes2011autologistic} focused on the methods about the estimation of the centered autologistic model. They studied maximum pseudo-likelihood estimation (PL) followed by parametric bootstrap, Monte Carlo maximum likelihood (MCML) and MCMC Bayesian approaches to inference and describe ways to optimize the efficiency of these algorithms. They also compared the performances of the three approaches in a thorough simulation study and found that  
pseudo-likelihood inference, which is much easier to understand and to implement than the MCML and Bayesian approaches, is both statistically and computationally efficient for data sets that are large enough to allow valid inference. A package for the free software R is proposed in the purpose to fit centered spatial models \cite{hughes2014ngspatial}.

\subsection{Spatio-temporal autologistic models} 
Now let us use $\mathbf{Z}_t$ to denote a random field indexed by time $t$. $Z_{it}$ for  $i=1,...,n$ and  $t \in \mathbb{Z}$ is a  random binary variable indexed by site $i$ and time $t$. And the covariates $\mathbf{X}$ are $p$-vectors indexed by $i$ and $t$.

Zhu {\it et al.} (2005) \cite{zhu2005modeling} generalized the autologistic regression models to account for covariates, spatial dependence, and temporal dependence simultaneously. The model  specifies the joint distribution of $\{ \mathbf{Z}_t: t \in \mathbb{Z}\}$ by a family of conditional distributions:

\begin{eqnarray}
\lefteqn{\mathbb{P}(\mathbf{Z}_{t_1},...,\mathbf{Z}_{t_2}|\mathbf{Z}_t; t \neq t_1,...,t_2)}    \label{ST1}\\
& \propto & exp \left\{ \sum_{t'=t_1}^{t_2}\left[\sum_{i=1}^{n}\mathbf{X}_{i,t'}^T\boldsymbol{\beta}Z_{i,t'}+\frac{1}{2} \sum_{i=1}^{n}\sum_{j\in N_i} \beta_{p+1}Z_{i,t'}Z_{j,t'}\right] 
+ \sum_{t'=t_1}^{t_2+1}\sum_{i=1}^{n} \beta_{p+2}Z_{i,t'}Z_{i,t'-1}\right\}. \nonumber
\end{eqnarray}

\noindent 
for all $t_1, t_2\in \mathbb{Z}^2$ such that $t_1<t_2$, where $\mathbf{X}_{i,t'}$ is the $k$-vector of the covariates at site $i$ and time $t$. Note that the specification is consistent for all $t_1<t_2$, and the joint distribution of $\{ \mathbf{Z}_t: t \in \mathbb{Z}\}$  can be shown to exist by Theorem 2.1.1 of \cite{guyon1995random}.
If  $N_{i,t}=\{(j,t): j\in N_i \} \cup \{(i,t-1), (i,t+1)\}$ denotes a neighborhood set for the $i$th site and the $t$th time point, the full conditional distribution of the model is   

$$[Z_{i,t}| Z_{i',t'}: (i',t') \neq (i,t)]=[Z_{i,t}|Z_{i',t'}: (i',t')\in N_{i,t}]  \quad \text{and }$$

\begin{eqnarray}
\lefteqn{\text{logit}(\mathbb{P}(Z_{i,t}=1|Z_{i',t'}: (i',t')\in N_{i,t}))} \label{ST2}\\
& = &\sum_{k=0}^p
\mathbf{X}_{i,t'}^T\boldsymbol{\beta}+\sum_{j\in N_i} \beta_{p+1}Z_{j,t'}+\beta_{p+2}(Z_{i,t-1}+Z_{i,t+1}).
\nonumber
\end{eqnarray}
\\
\cite{zheng2008markov} pointed out that one drawback of \cite{zhu2005modeling} is that parameter estimation for the spatio-temporal autologistic regression model was based on maximum pseudo-likelihood whose statistical efficiency is not well established. \cite{zheng2008markov} propose a fully Bayesian approach and compared it to the maximum pseudo-likelihood and MCMC maximum likelihood approaches. Another drawback is the difficulty to use this model for application. It seems unrealistic to express the probability of occurrence of an event according to the future.
It is probably why \cite{zhu2008autologistic} developed a spatio-temporal autologistic regression model which depends  only on the past. They drew statistical inference of this model via maximum likelihood estimation.
On one hand, they assume that the temporal dependencies satisfies the following property $[\mathbf{Z_t}|\mathbf{Z}_{t'}, t'= t-1, t-2,...]= [\mathbf{Z_t}|\mathbf{Z}_{t'},t'= t-1,..., t-S]$. On the other hand, for a given time $t$, the spatial field is a spatial Markov random field and the conditional probabilities are given by
\begin{eqnarray}
\lefteqn{\text{logit}( \mathbb{P}(Z_{it}=1|Z_{jt}, j\in N_i;\mathbf{Z}_{t'},t'= t-1,..., t-S) )}\\
& = &\mathbf{X}_{i,t'}^T\boldsymbol{\beta}+\sum_{j\in N_i} \beta_{p+1}Z_{j,t}+\sum_{s=1}^S \beta_{p+1+s} Z_{i,t-s}. \nonumber
\end{eqnarray}But nothing is said about the existence of the joint distribution of such a field in the paper.

To overcome the interpretation difficulties pointed by \cite{caragea2009autologistic} in the spatial case,  \cite{wang2013analysis} were the first to develop a centered parametrization version of (\autoref{ST2}) in a spatio-temporal framework. They propose the following modelling:  
\begin{eqnarray}
\lefteqn{ \text{logit}(\mathbb{P}(Z_{i,t}=1|Z_{i',t'}: (i',t')\in N_{i,t};\mathbf{X}))} \label{STcenter}\\
& = & \sum_{k=0}^p
\mathbf{X}_{i,t'}^T\boldsymbol{\beta}+\sum_{j\in N_i} \beta_{p+1}Z^{\ast}_{j,t}+\beta_{p+2}(Z^{\ast}_{i,t-1}+Z^{\ast}_{i,t+1}),
\nonumber
\end{eqnarray}

\begin{equation}
\text{with} \quad Z^{\ast}_{i,t}= Z_{i,t}- \frac{exp(\mathbf{X}_{i,t}^T\boldsymbol{\beta})}{1+exp(\mathbf{X}_{i,t'}^T\boldsymbol{\beta})}. \label{eq.centre1} 
\end{equation}

They proposed expectation-maximization algorithm to maximise the pseudo-likelihood and Monte Carlo expectation-maximization
likelihood, as well as consider Bayesian inference to obtain the estimates of model
parameters, and they found that Monte Carlo expectation-maximization
likelihood is optimal considering the computational cost and the estimation accuracy. Further, they compared the statistical efficiency of these approaches.

In the next section we propose a new centered spatio-temporal model and we believe that such model is better adapted for parameter interpretation in a spatio-temporal context.

\section{A new centered spatio-temporal autologistic model}
\subsection{Model specification}
In this section, we propose a new spatio-temporal autologistic model, specified by Markov field Markov chain \cite{gaetan2008modelisation} and accounting for spatio-temporal covariates. In order to overcome the problem of bias and interpretation of the instantaneous spatial effect, we propose to center the corresponding covariates term.  We define the model as follows. First, we assume that, conditionally to the covariates,  $\{ \mathbf{Z}_t, t=1,2,...\}$ is a Markov chain:

$$ [\mathbf{Z}_t| \mathbf{Z}_{t-1},\mathbf{Z}_{t-2}, ... ,\mathbf{X}]= [\mathbf{Z}_t| \mathbf{Z}_{t-1},\mathbf{X}_t]$$

\noindent where $\mathbf{X} $
is the spatio-temporal process of vector of the covariates $ (X_{i,t})_{1 \leq i \leq n , 1\leq t \leq T}$. Moreover, we assume that $\mathbf{Z}_t$ is a Markov random field conditional on $\mathbf{Z}_{t-1}$ with spatial neighbor structure $N_i$, that means 

$$ [Z_{i,t}| Z_{j,t}, j \neq i; \mathbf{Z}_{t-1},\mathbf{X}_t]= [Z_{i,t}| Z_{j,t}, j \in N_i, \mathbf{Z}_{t-1},\mathbf{X}_t].$$
More precisely, we define the conditional probabilities of $Z_{i,t}$ to be equal to one by:
\begin{equation}
\label{eq:model}
 \text{logit}(\mathbb{P}(Z_{i,t}=1| Z_{j,t}; j \neq i,\mathbf{Z}_{t-1},\mathbf{X}_t))=\mathbf{X}_{i,t}^T\boldsymbol{\beta}+    \rho_1\sum_{j\in N_i}Z^{\ast \ast}_{j,t}+ \rho_2 Z_{i,t-1},
\end{equation}
\begin{equation}
 \text{where} \quad   Z^{\ast\ast}_{i,t}=Z_{i,t}-\frac{exp(\mathbf{X}_{i,t}^T\boldsymbol{\beta}+\rho_2 Z_{i,t-1})}{1+exp(\mathbf{X}_{i,t}^T\boldsymbol{\beta}+\rho_2 Z_{i,t-1})}. \label{eq.centre2} 
\end{equation}

We will discuss this choice of centering in the next section but we can note that this new centered specification is similar with a hierarchical model with a latent auto-regressive model of order 1 given by:
$$\text{logit}(\mathbb{P}(\mathbf{Z}_{i,t}=1| \xi_{i,t}, \mathbf{X}_{i,t}))= \mathbf{X}_{i,t}^T\boldsymbol{\beta}+\xi_{i,t},$$

$$  \xi_{i,t}=\rho_2\xi_{i,t-1}+ \omega_{i,t},  \qquad \text{with following correspondences:}$$

$$ Z_{i,t-1} \approx \xi_{i,t-1} \quad \text{and}$$

$$ \sum_{j \in N_i} Z^{\ast\ast}_{j,t}= \sum_{j \in N_i}( Z_{j,t}-\frac{exp(\mathbf{X}_{j,t}^T\boldsymbol{\beta}+\rho_2 Z_{j,t-1})}{1+exp(\mathbf{X}_{j,t}^T\boldsymbol{\beta}+\rho_2 Z_{j,t-1})}) \approx  \omega_{i,t}.$$

\subsection{Model Interpretation}
There are two main differences between one-step centered of \cite{wang2013analysis} given by Equation (\ref{eq.centre1}) and what our new centering  given by Equation (\ref{eq.centre2}). The first one is that we do not center de temporal term $Z_{i,t-1}$. Indeed, there is no difficulty for the interpretation of $\rho_2$ because unlike the $\sum_{j\in N_i}Z_{j,t}$, $Z_{i,t-1}$ is known at time $t$ and can be treated like the other covariates. The second one is that the spatial neighborhoods $Z_{j,t}$'s are centered differently: the former is centered with $\frac{exp(\mathbf{X}_{j,t}^T\boldsymbol{\beta})}{1+exp(\mathbf{X}_{j,t}^T\boldsymbol{\beta})}$, and the latter is centered with $\frac{exp(\mathbf{X}_{j,t}^T\boldsymbol{\beta}+ \rho_2 Z_{j,t-1})}{1+exp(\mathbf{X}_{j,t}^T\boldsymbol{\beta}+ \rho_2 Z_{j,t-1})}$. It has to be noted that due to the link function, the expectation $E(Z_{it}|\mathbf{Z}_{t-1},\mathbf{X}_{t})$ equals $\frac{exp(\mathbf{X}_{i,t}^T\boldsymbol{\beta}+ \rho_2 Z_{i,t-1})}{1+exp(\mathbf{X}_{i,t}^T\boldsymbol{\beta}+ \rho_2 Z_{i,t-1})}$. 
The construction of this new centered autologistic model is explained as follows;
%
%
%
%
the key issue of this centering is to separate the large- and small-scale structures. The parameters of spatio-temporal dependence $\rho_1, \rho_2$ can be hierarchically interpreted as practical biological processes:
\begin{itemize}
	\item  Instantaneous spatial dependence $\rho_1$. To model illness, we expect some kind of common sensibility traducted by $\rho_1\geq 0$.\\
	Strong spatial dependence indicates a highly aggregated spatial structure, this gives a guide to identify and monitor the aggregated zones with high infection possibility.
	\item  Temporal dependence. Again, we expect that the illness likely remains at a place such that $0 \leq\rho_2$.  ( $\rho_2<0$ would indicate a temporal evolution with high frequency at 2-year cycle, this is not adapted for most of the biological processes); $\rho_2$ is a term of autoregressive regression.
	Strong temporal dependence can be interpreted as a smooth temporal evolution. If the exterior effects are stable, the individuals  have a tendency to keep their status. This may indicate no need to monitor two consecutive years if the exterior factors are similar between two years. 
	\end{itemize}
	
\subsection{Existence of the joint distribution}
It is probably not possible to show the existence of the distribution of the whole spatio-temporal processes defined by Equations (\ref{eq:model}) for all years $t$ and sites of the lattice $i$. In our case, the functionals involved in the right side of (\ref{eq:model}) are not invariant by permutating the temporal indices while it was the case in \cite{zhu2005modeling} given by equations (\ref{ST1}). To prove the existence of our process, we consider the framework of Markov chain of Markov fields presented in \cite{guyon2002markov}. Using the Hammersley-Clifford theorem given for instance in \cite{gaetan2008modelisation}, we can show the existence of the joint law of spatial process $\mathbf{Z}_t$ for a fixed $t$ given the past of $\mathbf{Z}_t$ and the current information about the covariates and derive an expression of this conditional joint spatial law. In this way the existence of the law of the whole spatio-temporal process, is trivial by recursivity. Moreover, we can also give the expression of the conditional transition probabilities of the spatial field from one year to the next one. 
We have the following result.
\begin{theorem}
\label{th:exist}
In the framework of spatio-temporal process given by (\ref{eq:model}), let us denote  $\cF^X_t = \sigma\{\mathbf{X}_{i,s}, 1 \leq i \leq n, s \leq t\}$ and $\cF^Z_t = \sigma\{{Z}_{i,s}, 1 \leq i \leq n, s \leq t\}$ the $\sigma$-algebra generated by the covariates and the process of interest respectively.

 Given $\cF^Z_{t-1} \wedge \cF^X_{t}$ the conditional joint law of  $\mathbf{Z}_t$ denoted by $\pi_t(. \mid \cF^X_t, \cF^Z_{t-1})$  is well defined. Moreover, for $\mathbf{z} = (z_1, \ldots, z_n) \in \{0,1\}^n$,  the spatial conditional joint law is of the form
\begin{eqnarray*}
 \pi_t(\mathbf{z}\mid \cF^X_t, \cF^Z_{t-1})&=& C(\cF^X_t, \cF^Z_{t-1}) \exp(\sum_{i \in S}\Phi_i(z_i) + \sum_{\{i,j\}}\Phi_i(z_i,z_j)) \quad \text{with}\\
\Phi_i(z_i, \cF^X_t, \cF^Z_{t-1})  &=& z_i \left( \mathbf{X}_{i,t}^T\boldsymbol{\beta}-  \rho_1\sum_{j\in N_i}\frac{exp(\mathbf{X}_{j,t}^T\boldsymbol{\beta}+\rho_2 Z_{j,t-1})}{1 + exp(\mathbf{X}_{j,t}^T\boldsymbol{\beta}+\rho_2 Z_{j,t-1})}+ \rho_2 Z_{i,t-1}\right)\\
\Phi_i(z_i, z_j)   &=&{\rho_1 \inds{\{j \in N_i\}}}z_iz_j
\end{eqnarray*}and $C(\cF^X_t, \cF^Z_{t-1})$ can be considered as constant if the past of process $\mathbf{Z}_t$ and the current values at time $t$ of the covariates are known.\\

The transition probabilities of the Markov chain are 
\begin{eqnarray*}
 \mathbbm{P}(\mathbf{y},\mathbf{z}\mid \cF^X_t)&=& C(y, \cF^X_t){\exp\left(\sum_{i \in S} (z_i\mathbf{X}_{i,t}^T\boldsymbol{\beta} +\rho_1 \sum_{j \in N_i}z_iz_j )\right)}
\\
& \times & {\exp\left( \sum_{i \in S}z_i(\rho_2y_i -  \rho_1\sum_{j\in N_i}\frac{exp(\mathbf{X}_{j,t}^T\boldsymbol{\beta}+\rho_2 y_j)}{1 + exp(\mathbf{X}_{j,t}^T\boldsymbol{\beta}+\rho_2 y_j)}- \rho_2 y_i)\right)}\\
\end{eqnarray*}
\end{theorem}

\noindent 
{\bf Proof: } We use Hammersley-Clifford theorem given for instance in \cite{gaetan2008modelisation} in a conditional form. With the above notation, let us define two assumptions given by (\ref{expo3}) and (\ref{cond13}) that says that if 
\begin{eqnarray}
\label{expo3}
\lefteqn{\text{logit}(\mathbb{P}(Z_{i,t}=z_i  | Z^i_t=z^i,{\cF^X_t, \cF^Z_{t-1} }))}\\
 & = & A_i(z^i, {\cF^X_t, \cF^Z_{t-1} } )B_i(z_i, {\cF^X_t, \cF^Z_{t-1} }) + C_i(z_i,{\cF^X_t, \cF^Z_{t-1} }) + D_i(z^i, {\cF^X_t, \cF^Z_{t-1} } )  \nonumber
  \end{eqnarray}

and  if  $\forall i \neq j$, it exists $\alpha_i$ and $\beta_{ij}= \beta_{ji}$ such that
\begin{equation}
\label{cond13}
A_i(z^i, {\cF^X_t, \cF^Z_{t-1} })= \alpha_i({\cF^X_t, \cF^Z_{t-1} }) + \sum_{j \neq i} \beta_{ij} B_j(z_j,{\cF^X_t, \cF^Z_{t-1} )}
\end{equation}
 The conditional laws satisfying (\ref{expo3}) (\ref{cond13}) are consistent to a joint distribution which is a Markov random field. If we rewrite (\ref{eq:model}) in the following form, it is easy to see that our model satisfies the required conditions.
  \begin{eqnarray*}
  \lefteqn{\text{logit}(\mathbb{P}(Z_{i,t}=1| Z_{j,t}; j \neq i,\mathbf{Z}_{t-1},\mathbf{X}_t))}\\
   &= & \underbrace{\mathbf{X}_{i,t}^T\boldsymbol{\beta}-  \rho_1\sum_{j\in N_i}\frac{exp(\mathbf{X}_{j,t}^T\boldsymbol{\beta}+\rho_2 Z_{j,t-1})}{1+exp(\mathbf{X}_{j,t}^T\boldsymbol{\beta}+\rho_2 Z_{j,t-1})})+ \rho_2 Z_{i,t-1}}_{=\alpha_i(\cF^X_t, \cF^Z_{t-1}))}+ \sum_{j \neq i} \underbrace{\rho_1 \inds{j \in N_i}}_{\beta_{ij}}\underbrace{Z_{j,t}}_{B_j(z_j)}.
\end{eqnarray*}
The expression of $\pi$ comes directly from Hammersley-Clifford theorem and the transitions probabilities from \cite{guyon2002markov}.
\hspace{\stretch{1}}$ \Box$

\medskip
\noindent

\section{Comparative Simulation}
\subsection{Simulation objective}

The idea of proposing the new centered model is to make an agreement between large-scale model and data structure. Thus we compare three models: the model without centering that Caragea and Kaiser called "traditional" and one-step centered  as well as the new centered models, defined again below. We want to verify if the marginal structure of data reflects the large-scale structure. The three compared models are given by the general following equation differing according to the kind of centering of $Z^{\ast\ast}_{i,t}$.

\begin{equation*}
\text{logit}(p_{i,t})=\mathbf{X}_{i,t}^T\boldsymbol{\beta}+    \rho_1\sum_{j\in N_i}Z^{\ast \ast}_{j,t}+ \rho_2 Z_{i,t-1},
\end{equation*}
\begin{eqnarray}
  Z^{\ast\ast}_{i,t} &= & Z_{i,t} \qquad \text{traditional model} \label{eq.trad} \\
    Z^{\ast\ast}_{i,t}& = & Z_{i,t}-\frac{exp(\mathbf{X}_{i,t}^T\boldsymbol{\beta})}{1+exp(\mathbf{X}_{i,t}^T\boldsymbol{\beta})} \qquad \text{one-step model} \label{eq.onestep} \\
      Z^{\ast\ast}_{i,t}& = &Z_{i,t}-\frac{exp(\mathbf{X}_{i,t}^T\boldsymbol{\beta}+\rho_2 Z_{i,t-1})}{1+exp(\mathbf{X}_{i,t}^T\boldsymbol{\beta}+\rho_2 Z_{i,t-1})} \qquad \text{new model}  \label{eq.new} 
\end{eqnarray}

The agreement between spatial large-scale model structures and marginal data structures have been already examined by  \cite{caragea2009autologistic} for both centered and traditional spatial autologistic regression models. They showed that the  data structure of traditional autologistic regression model can not reflect the large-scale structure, and this difficulty can be alleviated by centered parametrization. \\

In this paper, we focus on examining the time variation of the large-scale model structure and  marginal data structure for the three  spatio-temporal autologistic models mentioned above in the case when the covariates depend only on the time $t$ but not on the location on the grid. Note that \cite{caragea2009autologistic} has already performed simulations to study the agreement for the spatial model and spatial data structures according to the different values of the spatial covariates.\\

Therefore, to simulate a dynamic process of Markov random field, we consider a temporal large-scale structure with a temporal tendency without spatial covariates. For site $i$ at year $t$,   we define a large model structure with one temporal covariate: $A_{it}=\beta_0+\beta_1X_{it}+ \rho_1\sum_{j\in N_i}Z^{\ast \ast}_{j,t}+ \rho_2 Z_{i,t-1}$, where $X_{it}=X_t$ is a temporal covariate, that is spatial constant at year t. Thus the average large-scale model at year $t$ is: 
$$ L_t=\frac{1}{n} \sum_{i=1}^n \frac{exp(\beta_0+\beta_1 X_t)}{1+exp(\beta_0+\beta_1 X_t)}= \frac{exp(\beta_0+\beta_1 X_t)}{1+exp(\beta_0+\beta_1 X_t)}.$$Moreover, we can also compute an average scale conditional to the past defined by:
$$ C_t=\frac{1}{n} \sum_{i=1}^n \frac{exp(\beta_0+\beta_1 X_t+ \rho_2 Z_{i, t-1})}{1+exp(\beta_0+\beta_1 X_t + \rho_2 Z_{i, t-1}))}.$$Note that this process is not deterministic but specific at each realization of the field $\mathbf{Z}_{t-1}$.

To represent marginal data structure, the marginal empirical data mean of the Markov random field at year t is computed from a simulated field at time $t$ as:

$$ D_t = \frac{1}{n} \sum_{i=1}^n Z^{sim}_{it}.$$

The objective of the simulation studies presented here is to compare the behaviour of $L_t$, $C_t$  and that of  $D_t$  for the three different models differing by the kind of centering of the spatial autoregression. 
\subsection{Sampling Algorithms}
\label{ss:SA}
\cite{hughes2011autologistic} proposed to use perfect sampling to generate Markov Random Field (MRF) samples. The advantage of the perfect sampling compared to Gibbs sampling is that we do not need the burn-in step, neither to decide the spacing numbers. It gives us the exact draws from a given target distribution when the algorithm completes.   However, its algorithm running time is random even is still finite. We do not know at which moment the lower chain and the upper chain coalesce. So the number of repetition is random. In our case,  we have to generate a Markov chain Markov random field, the simulation time of such a algorithm is  quite difficult to control.

Here we use Gibbs sampler but start at a perfect simulation sample, we call it PGS sampling here. It is less time consuming than perfect sampling, and do not need to decide burn-in and spacing when compared with Gibbs sampling. The PGS sampling was often used to generate the  simulations of autologistic model \cite{zhu2008autologistic,zheng2008markov,wang2013analysis}. 


\subsection{Simulations}

We focus on two types of large-scale model structures, the first one without covariate and the second one with  increasing temporal tendency given by a covariate depending only on the time. The two models are described in the two following sections \ref{ss:1} and \ref{ss:2} but they were simulated on a lattice at $20 \times 20$ sizes over $50$ time's units.  These model structures have been simulated with different values of the "auto-regression parameters" $(\rho_1, \rho_2)$, in order to evaluate the joint effects of $(\rho_1, \rho_2)$ to the agreement between large-scale structure models and data structures. 
One trajectory of each of the three models is drawn for each configuration of $(\rho_1, \rho_2)$.  To study the dispersion of the empirical large scale structure and confirm the possible tendencies  exhibited by the trajectories, we have performed $100$ independent realizations of process $D_t$  for each of the three models (traditional, one-step and new centered) and specified by different values of $(\rho_1, \rho_2)$ and then computed and drawn the empirical fluctuation dynamical intervals.  The simulations are presented in the following sections.

\subsubsection{Model 1}
\label{ss:1}
We first consider a model without covariates (that is with only an intercept). It is given by  $\text{logit}(p_{it})=\beta_0+ \rho_1\sum_{j\in N_i}Z^{\ast \ast}_{j,t}+ \rho_2 Z_{i,t-1}$, there is no temporal covariate in this model apart the term of temporal auto-regression. The baseline level of infection is chosen via $\beta_0$ such that $\frac{exp(\beta_0)}{1+exp(\beta_0)}=0.2$ and the initial field is generated by independent Bernoulli variables with parameter $p=0.2$. To see the effect of the two parameters $(\rho_1, \rho_2)$ on the difference between the models  a given value of parameter, we perform simulations and draw one trajectory from each model with different values of parameters $(\rho_1, \rho_2) \in \{0.3, 0.5, 0.7\}^2$. The empirical confidence intervals were computed with 100 realizations of independent trajectories for each models and  for different values of parameters $(\rho_1, \rho_2) \in \{0.5, 0.7\}^2$. 
%
%
%

\begin{figure}%
\hspace{-1cm}
\includegraphics[width=1\textwidth]{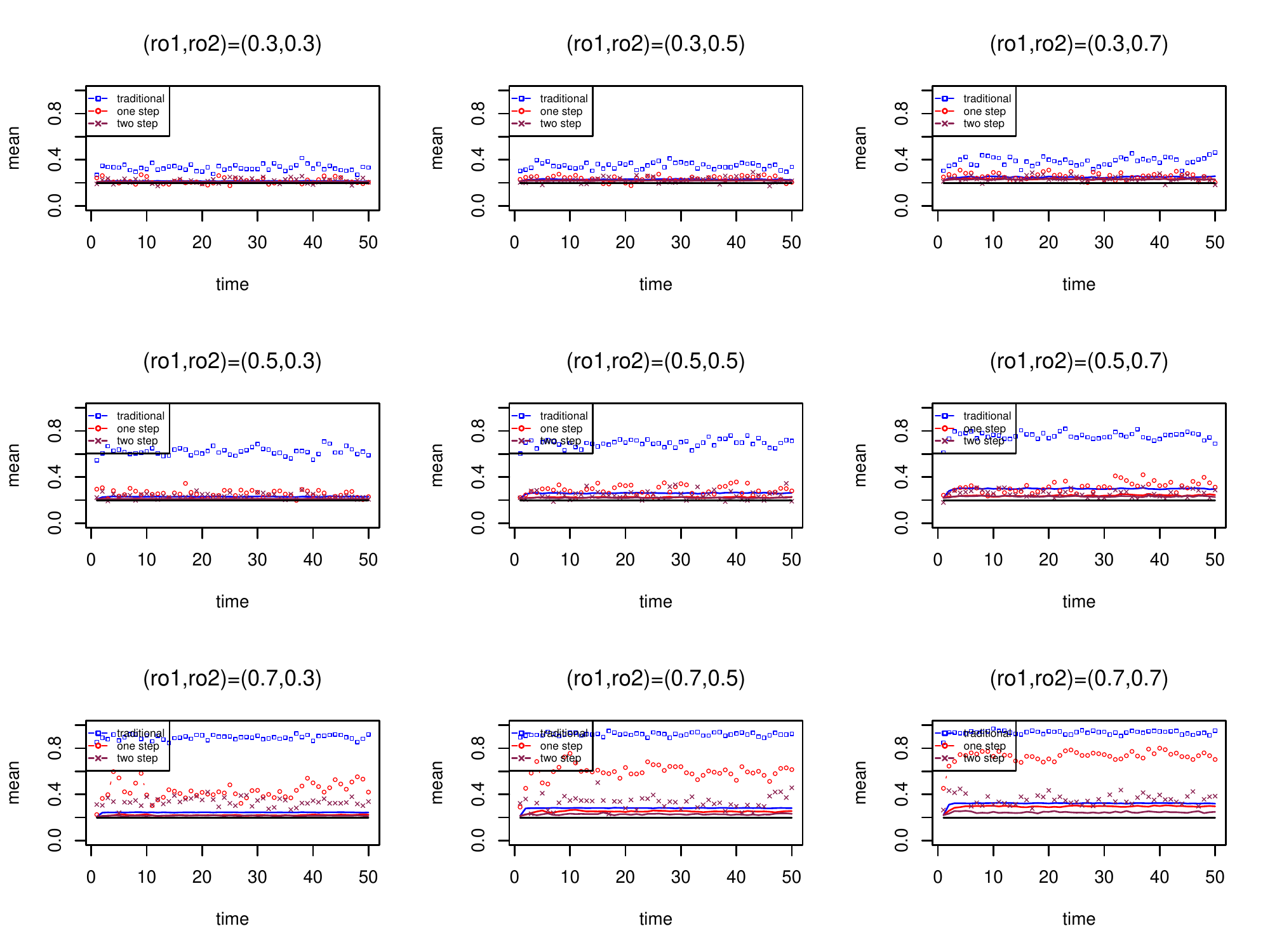}%
\caption{Comparison between large-scale model structure $L_t$ (represented by black line), the expected means according to the past (lines) and empirical mean of data structures $D_t$ (dots) for the traditional (blue box), the one-step(red circle)  and new (violet cross) centered models for different values of auto-regression parameters $(\rho_1, \rho_2)$. The grid is $20 \times 20$, and $0 \leq t \leq 50$, baseline infection $\frac{exp(\beta_0)}{1+exp(\beta_0)}=0.2$.}
\label{Interceptonly}%
\end{figure}

\begin{figure}%
\includegraphics[width=0.5\textwidth]{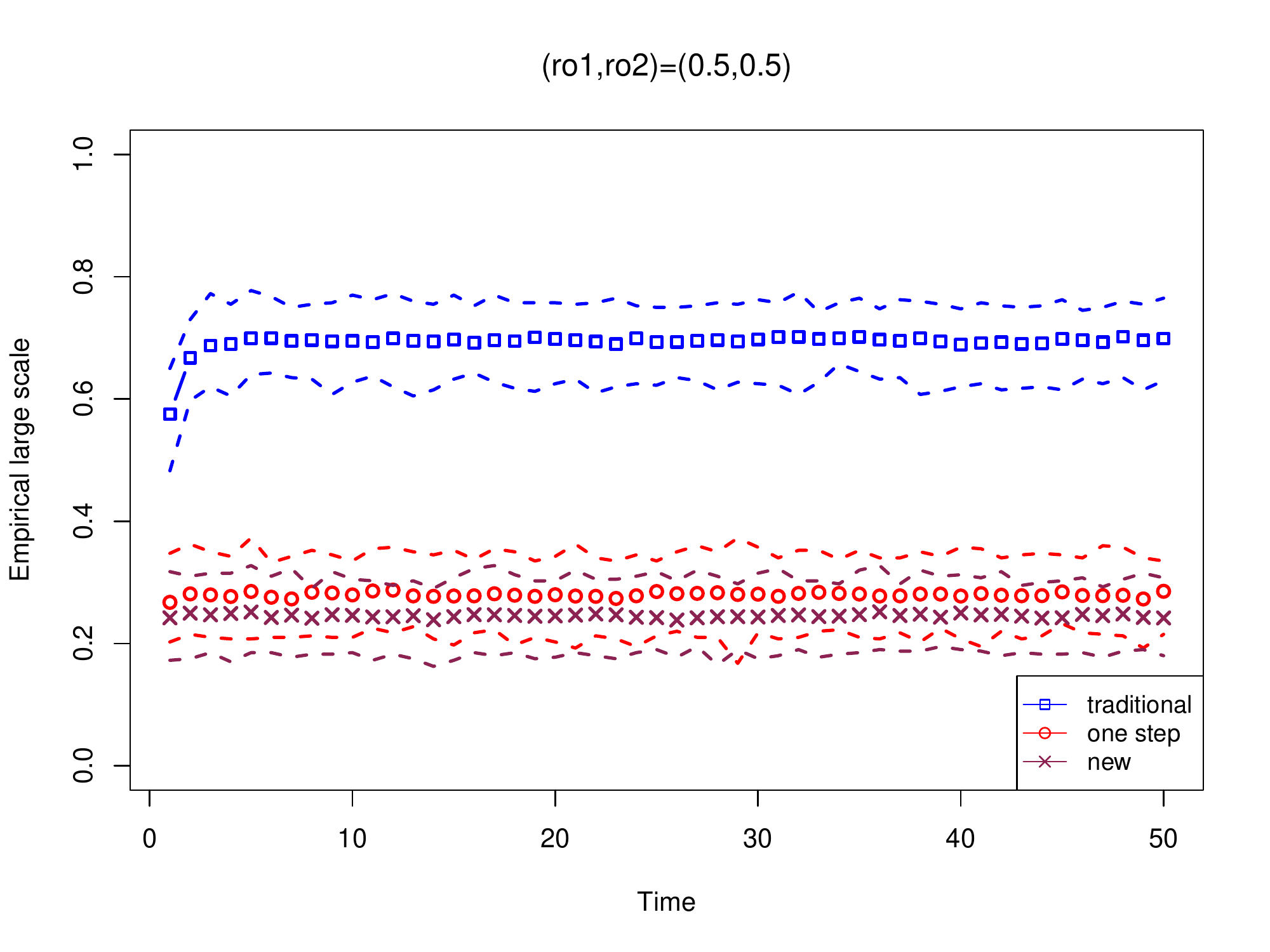}%
\includegraphics[width=0.5\textwidth]{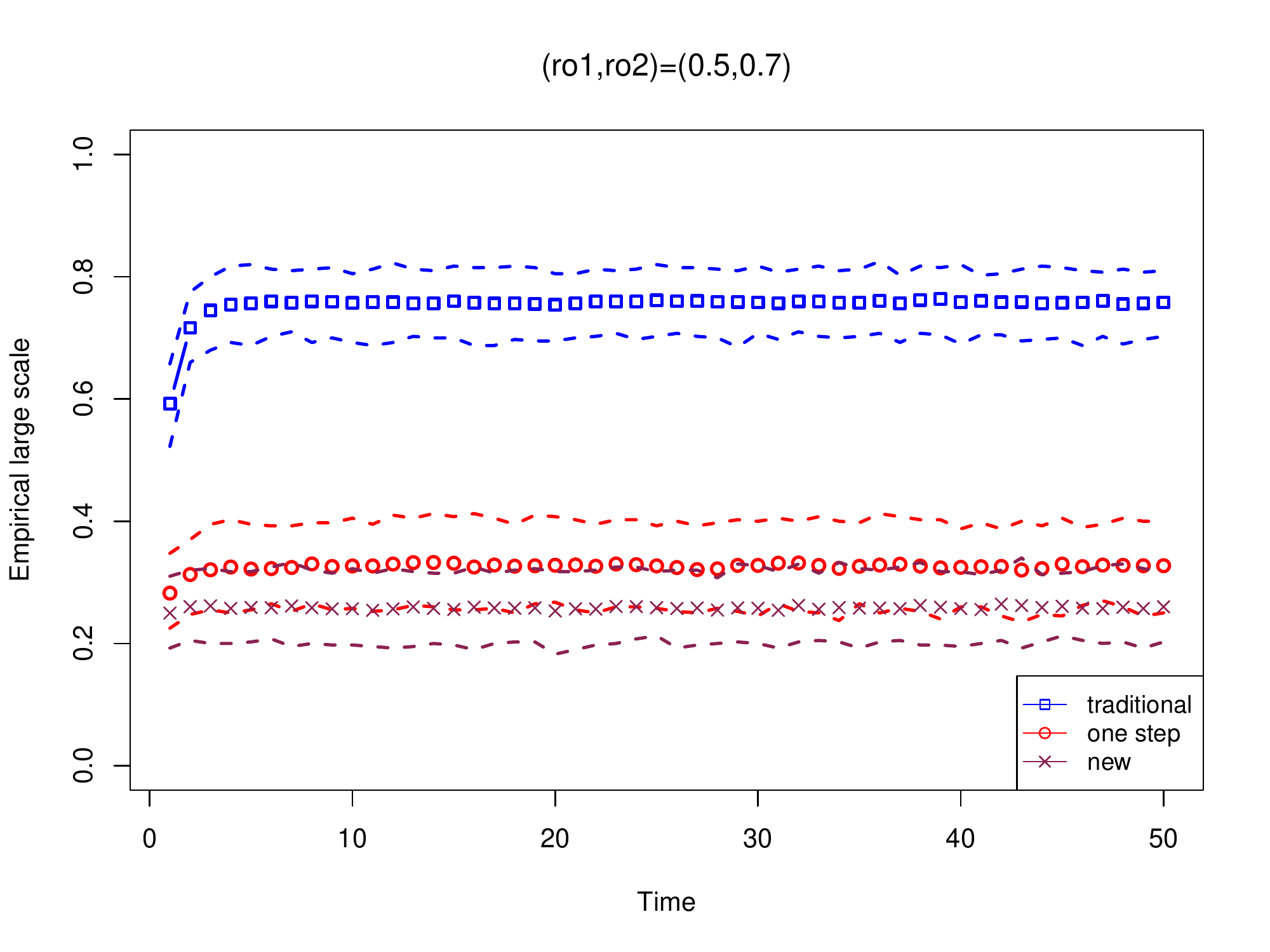}\\
\includegraphics[width=0.5\textwidth]{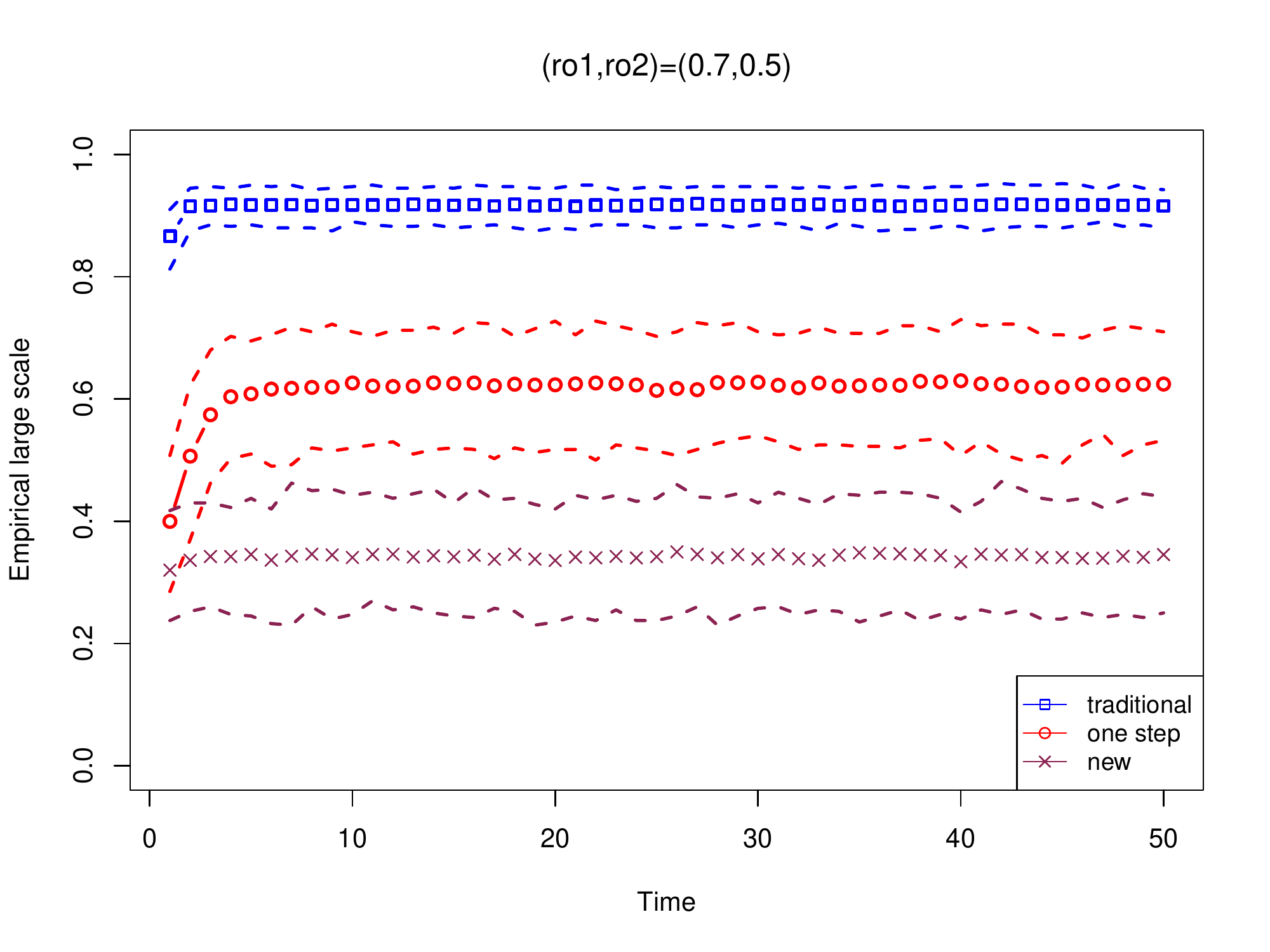}%
\includegraphics[width=0.5\textwidth]{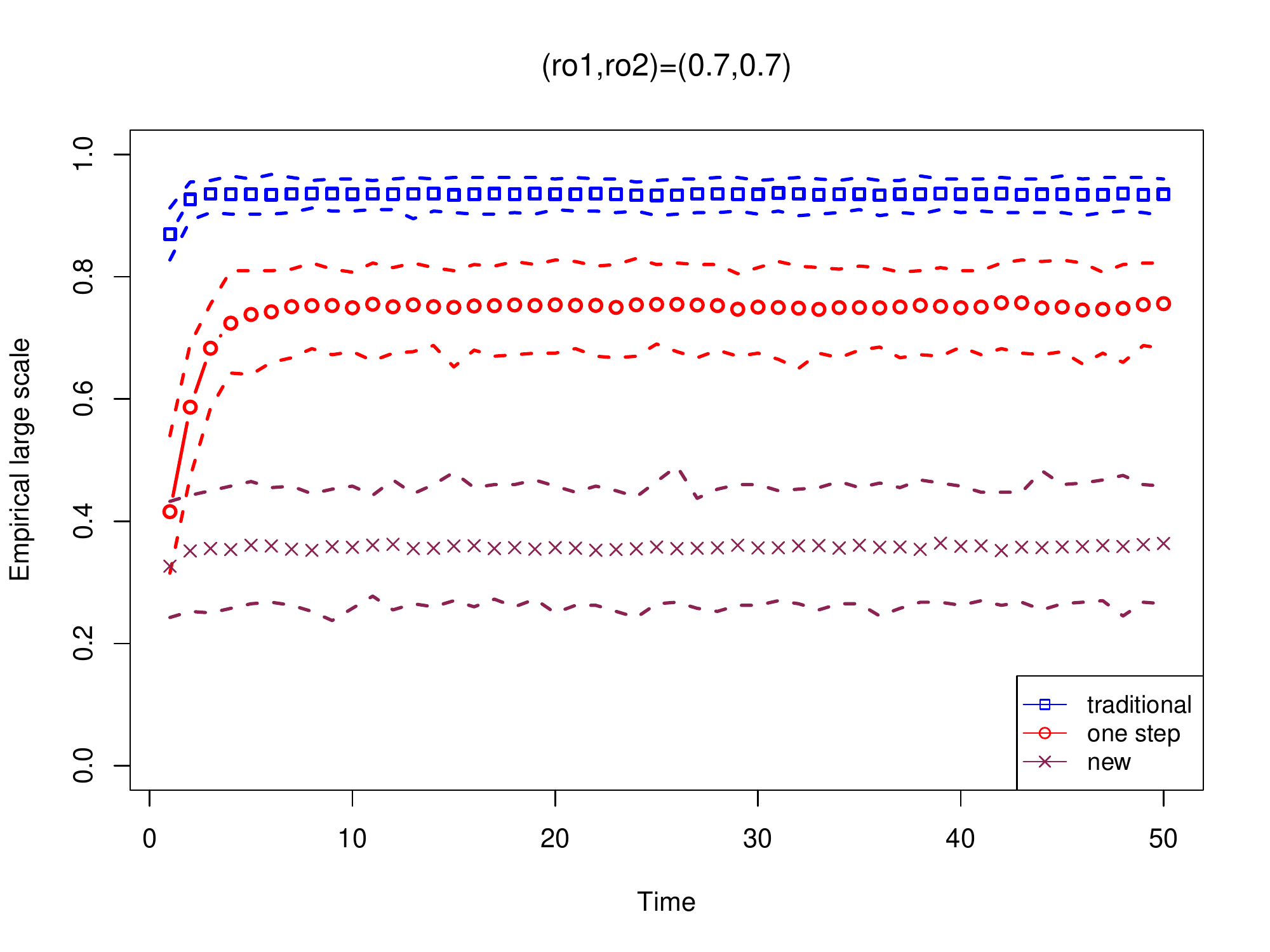}
\caption{Empirical confidence curves for the large scale structures $D_t$ for the model without covariate. The grid is $20 \times 20$, $t=50$, baseline infection $\frac{exp(\beta_0)}{1+exp(\beta_0)}=0.2$. About the centering, traditional  (resp. one-step and new centered) is drawn in blue box, (resp. red circle and violet cross).}
\label{rep1}%
\end{figure}

\subsubsection{Model 2}
\label{ss:2}
We concider here a model with a temporal trend. We consider  large-scale structures with one temporal covariate: $\beta_0+\beta_1X_t$. A said above, $X_t$ is constant spatially for each year but show  monotonic increasing with time: $X_t=t$.

We choose  $\beta_0$ such that $\frac{exp(\beta_0)}{1+exp(\beta_0)}=0.1$  and the initial field is generated by independent Bernoulli variables with parameter $p=0.1$. With this model, we have $L_t=\frac{exp(\beta_0+\beta_1 t)}{1+exp(\beta_0+\beta_1t)}$, a monotonic increasing function. With the chosen coefficients, $L_t$ increases from $0.1$ at time one to $0.94$ at time $50$. Again, for a trajectory, we compute the conditional mean given the past. Unlike the model without covariates, we expect an  empirical large-scale structure that increases with time.

\begin{figure}%
\hspace{-1cm}
\includegraphics[width=1\textwidth]{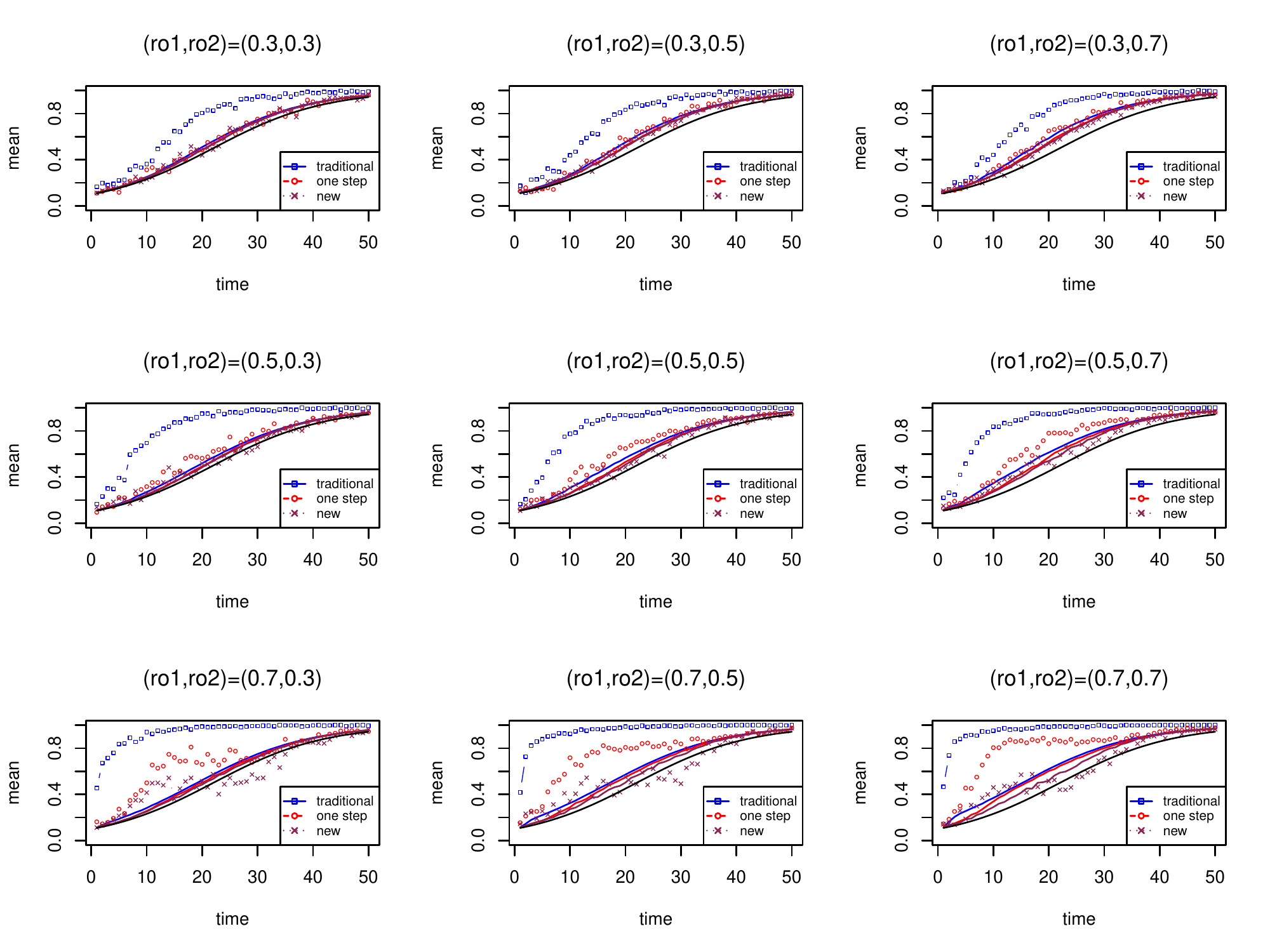}%
\caption{Comparison between large-scale model structure $L_t$ (represented by black line), the expected means according to the past (lines) and empirical mean of data structures $D_t$ (dots) for the traditional (blue box), the one-step(red circle)  and new  (violet cross) centered models for different values of auto-regression parameters $(\rho_1, \rho_2)$. The grid is $20 \times 20$, and $0 \leq t \leq 50$, baseline infection $\frac{exp(\beta_0)}{1+exp(\beta_0)}=0.1$, and covariate coefficient $\beta_1 =0$, and covariate $X_t = t$. }
\label{Increase}%
\end{figure}

\begin{figure}%
\includegraphics[width=0.5\textwidth]{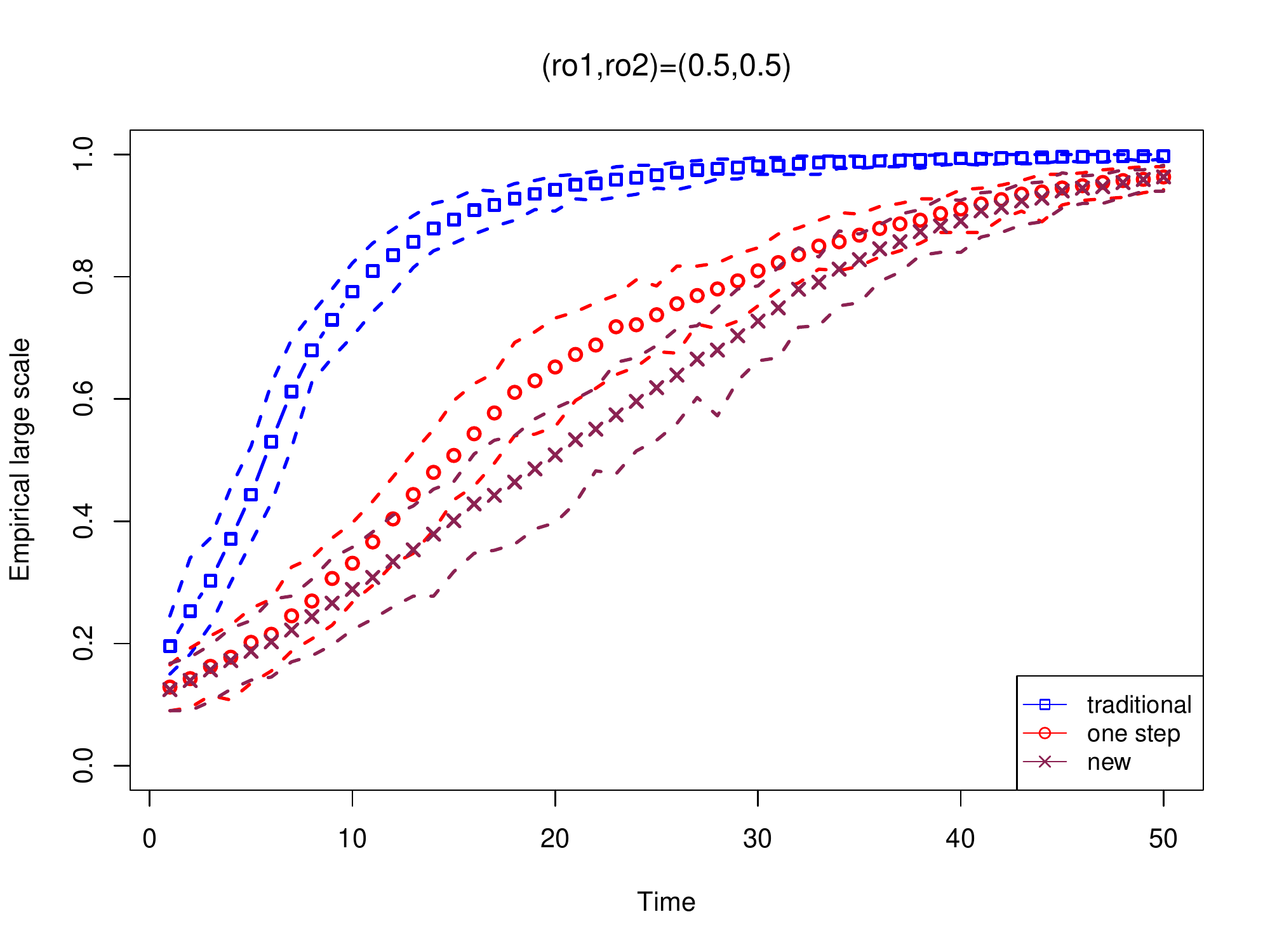}%
\includegraphics[width=0.5\textwidth]{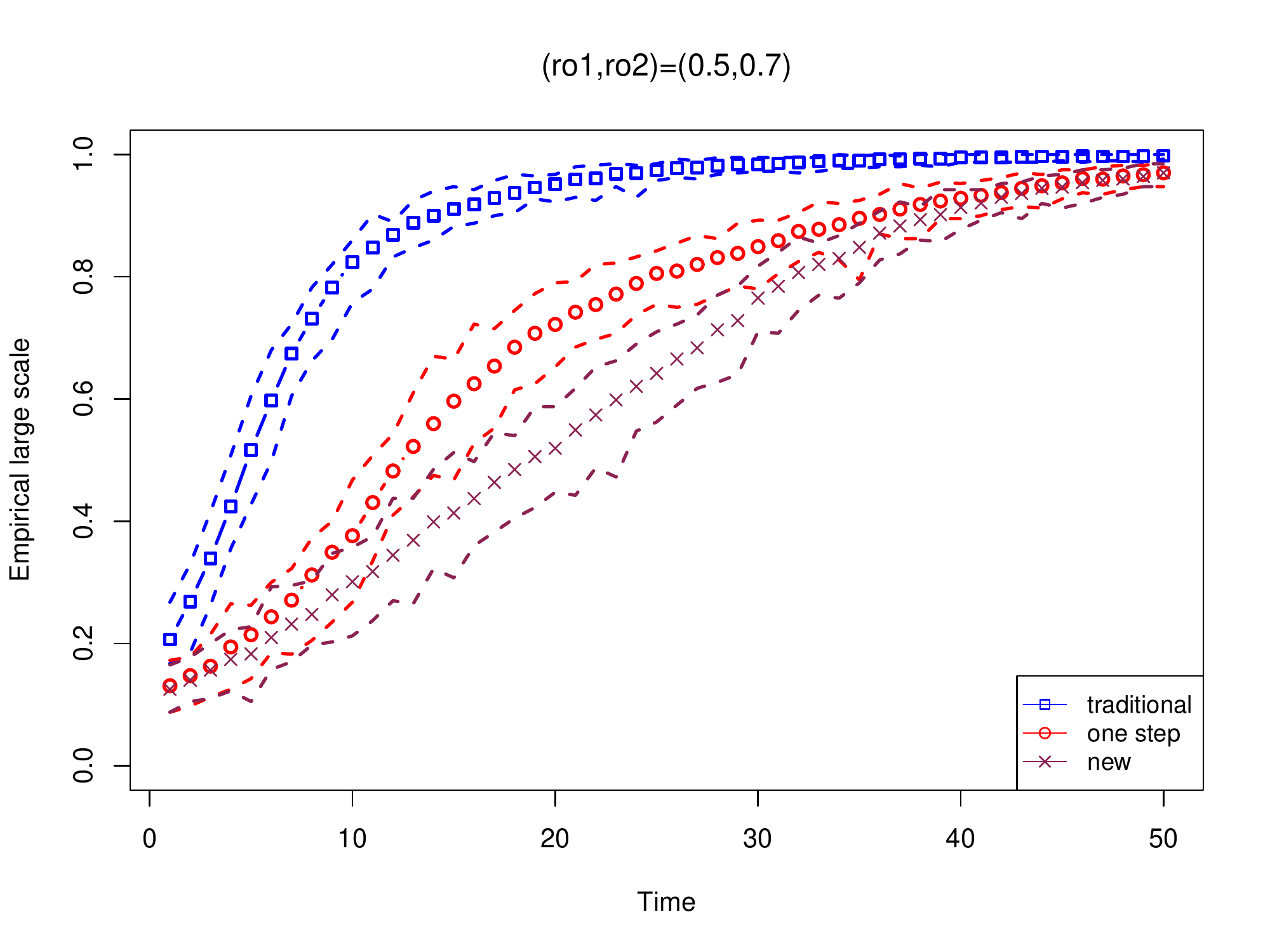}\\
\includegraphics[width=0.5\textwidth]{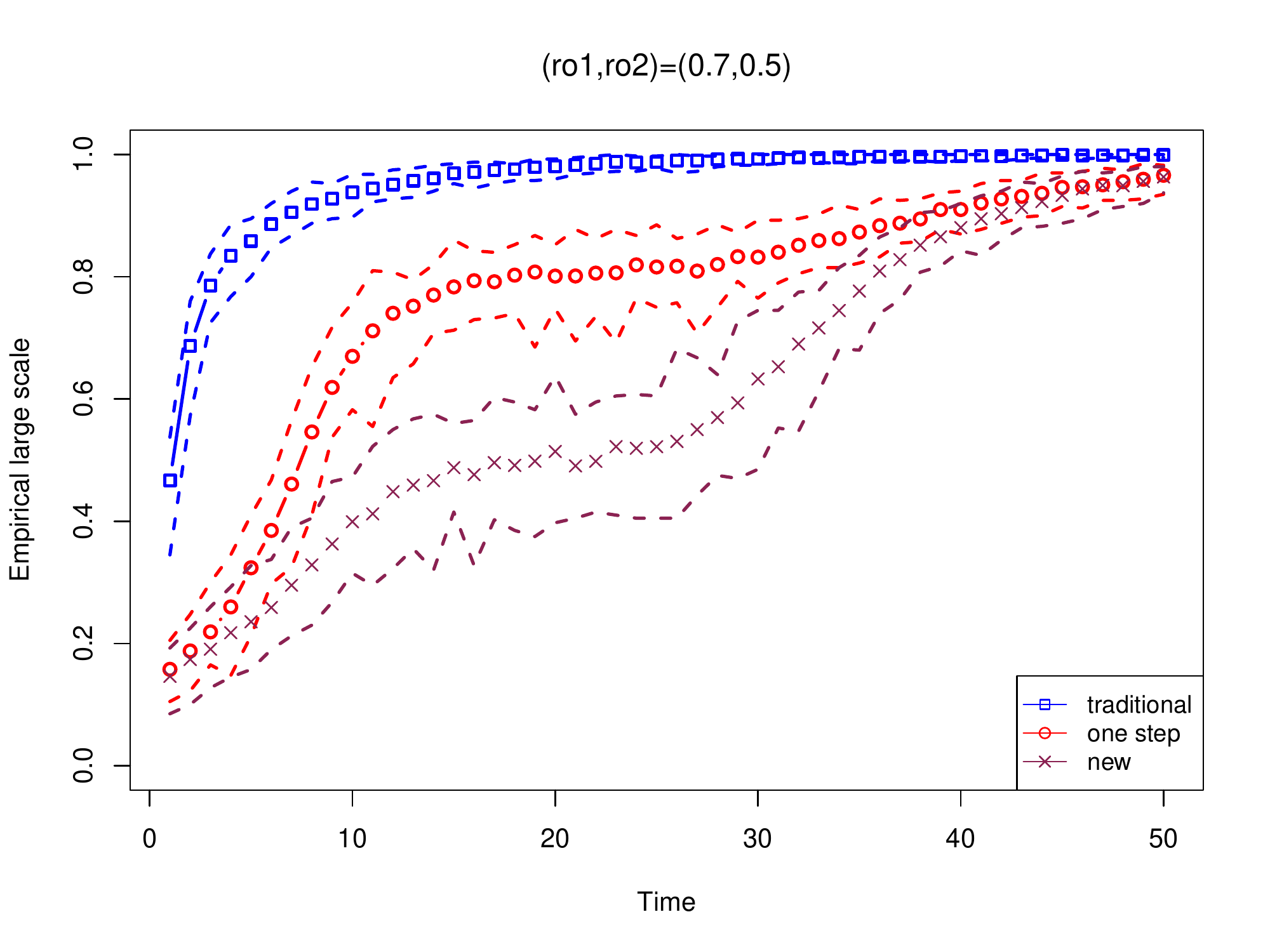}%
\includegraphics[width=0.5\textwidth]{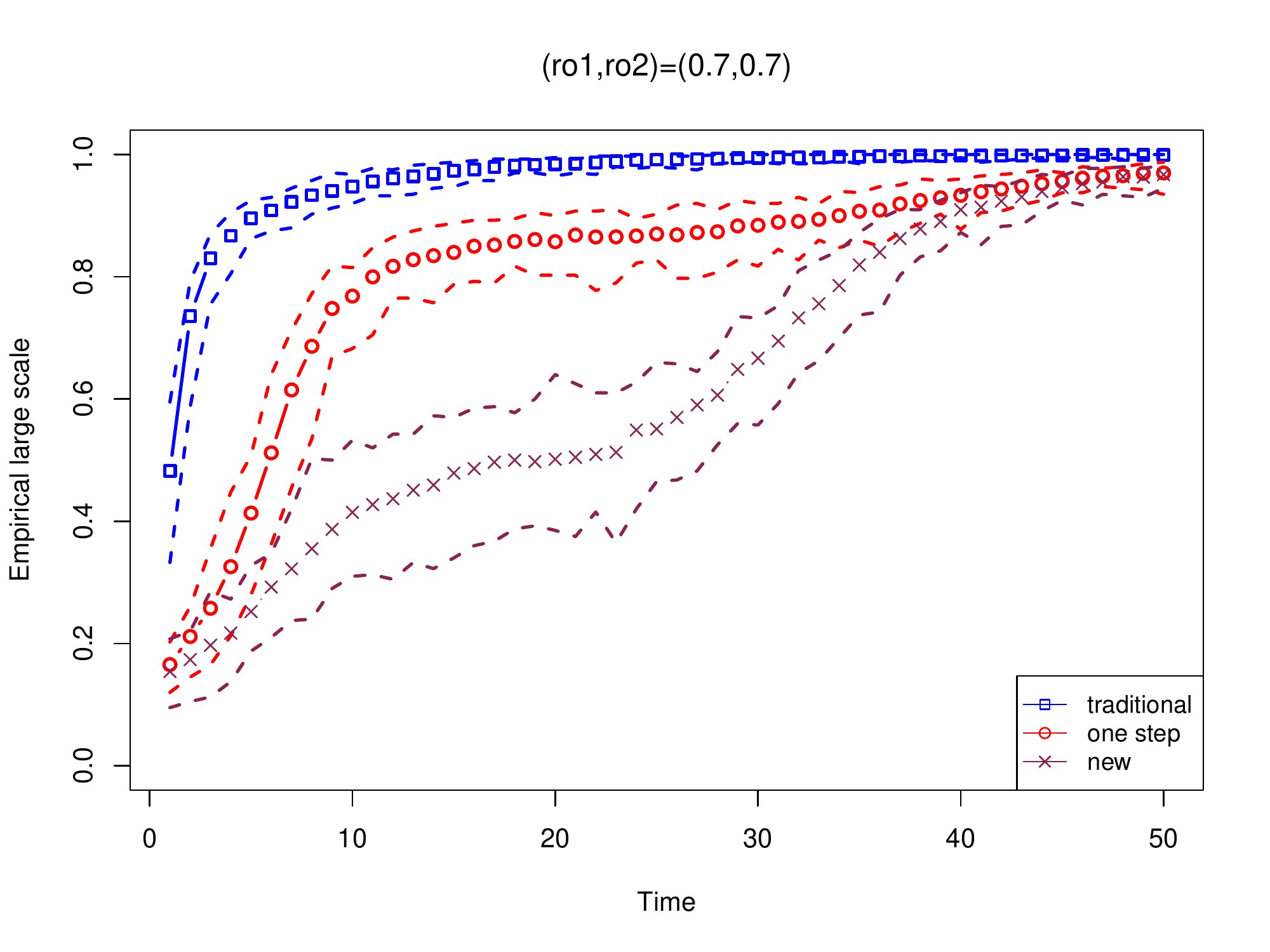}
\caption{Empirical confidence curves for the large scale structures $D_t$ for the model with temporal covariate $X_t=t$. The grid is $20 \times 20$, $t=50$,baseline infection $\frac{exp(\beta_0)}{1+exp(\beta_0)}=0.1$, and covariate coefficient $\beta_1 =0$, and covariate $X_t = t$. About the centering, traditional  (resp. one-step and new centere d) is drawn in blue box, (resp. red circle  and violet cross).}
\label{rep2}%
\end{figure}

\subsubsection{Results}
From \autoref{rep1} and \autoref{rep2}, we see that the spread of the realizations of the empirical mean $D_t$ is not very large and that for the given value of the parameter $(\rho_1, \rho_2)$, the models can differ more or less significantly or not.  We see that scatter of the empirical average of the spatial field for each $t$ is low enough to trust and interpret the difference between the individual curves shown in \autoref{Interceptonly}.

In \autoref{Interceptonly} (resp. \autoref{Increase}) we draw one trajectory from the each model without covariate (resp. with covariates) for different values of $\rho_1, \rho_2$  to study the difference between the models according to theses parameters. We see that the empirical mean $D_t$ for the traditional model is always greater than the expected large scale structure and than the realization of $D_t$ for the one and new centered models. The value of $D_t$ for the traditional model is not sensible to the value of the temporal autoregression parameter $ \rho_2$  but it is very sensible to the spatial auto-regression parameter $\rho_1$ and it increases as $\rho_1$ increases.

About the centered models, they show the same large scale  behavior when parameter $\rho_1$ equals $0.3$ or $0.5$. For $\rho_1= 0.7$, the two centered models show different large scale behavior but the $D_t$ for new centered model agrees with the expected mean behavior. The difference between them increases as $\rho_2$ increases.

In summary, the difference between the one and new centered autologistic model for the simulation are small except when both spatial and temporal dependence are relatively strong. The data structure of traditional or one-step centered model more over-reflects the large-scale structure and more seriously, it reflects a increasing trend which wrongly differs from the expected model structure and may confuse the further analysis.

\section{Estimation}

From now on,  we only consider the new centered model. We propose here a method of estimation by pseudo-likelihood maximisation and a method to select the most likely neighborhood structure of the data set. We show their performances by simulations. 

\subsection{Pseudo likelihood Estimation}
\label{sec:EMPL}
Since the structure of the new centered autologistic model is more complicated than the traditional or the one-step centered one, both Monte Carlo maximum likelihood (MCML) and Bayesian methods can be very heavy and sophisticated. We propose to estimate such models using Maximum pseudo-likelihood which is very easy to perform. Some authors have studied  the mathematical properties such as convergence of this estimator under the constraint of homogeneity and ergodicity of the Markov Random Field Markov Chain and other required assumptions can be found in \cite{guyon2002markov}. The imbrication of the parameters in the definition of the centered variables $Z^{\ast\ast}_{i,t}$'s and the presence of covariates make these conditions hard to verify, that is why, in the following, we only look at the empirical behaviour of this estimator. The pseudo-likelihood for our  model is given by the following formula.
\begin{equation}
\label{PL}
{\cal L} = \prod_{t=1}^T\prod_{1\leq i \leq n} p_{it} = \prod_{t=1}^T\prod_{1\leq i \leq n} \frac{\exp(\mathbf{X}_{i,t}^T\boldsymbol{\beta}+\rho_1\sum_{j \in N_i}Z^{\ast\ast}_{j,t} + \rho_2 Z_{i,t-1})}{1 + \exp(\mathbf{X}_{i,t}^T\boldsymbol{\beta}+\rho_1\sum_{j \in N_i}Z^{\ast\ast}_{j,t} + \rho_2 Z_{i,t-1})}.
\end{equation}with $Z^{\ast\ast}_{j,t}$ defined by (\ref{eq.new}).
Estimator is the vector of parameters $\boldsymbol{\theta}=\{\boldsymbol{\beta}, \rho_1, \rho_2\}$ that maximizes \autoref{PL}.
We see that the spatial auto-regression covariate $Z^{\ast\ast}_{j,t}$ imbricated the couple of parameters $(\rho_1, \rho_2)$ themselves so that it is not possible to consider $Z^{\ast\ast}_{j,t}$ as a common "external covariate". For this reason, the maximisation has to be made by an Expectation-Maximisation algorithm. We give the details of this algorithm below. 

\subsubsection{Algorithm}
Let us denote $\boldsymbol{\theta}=(\boldsymbol{\beta}, \rho_1, \rho_2)$, the parameters to estimate. We applied the EMPL (Expectation maximization pseudo-likelihood), the principle is the same as described in \cite{zheng2008markov}, but with two iteration steps to accelerate the numerical algorithm/calculation. \\

The principles are  as follows,
\begin{itemize}
	\item Initialisation: to obtain the estimation of $\boldsymbol{\theta_1}=(\boldsymbol{\beta}, \rho_2)$, denoted by $(\tilde{\boldsymbol{\beta}},\tilde{\rho_2})$, from model $logit(p_{it})=\mathbf{X}_{i,t}^T\boldsymbol{\beta}+\rho_2 Z_{i,t-1}$, we maximise the corresponding  log pseudo(partial)-likelihood by Quasi-Newton.
	\item Step 2: to obtain the estimation of $\boldsymbol{\theta}=(\boldsymbol{\beta}, \rho_1, \rho_2)$, denoted by $\breve{\boldsymbol{\theta}}= (\breve{\boldsymbol{\beta}},\breve{\rho_1}, \breve{\rho_2} )$, for the new centered autologistic model:
	\begin{enumerate}
	\item Initialization: Set initial values: $\boldsymbol{\theta}^{0}=( \tilde{\boldsymbol{\beta}}, 1, \tilde{\rho_2} )$
	\item Expectation: Given $\boldsymbol{\theta}^{l-1}$, compute the $Z^{\ast \ast}_{j,t}$'s by removing the corresponded trend. 
	\item Maximization: Obtain $\boldsymbol{\theta}^{l}$ by maximizing the log pseudo-likelihood 
	by Quasi-Newton.
	\end{enumerate}	
	\item  Obtain estimates $( \breve{\boldsymbol{\beta}},\breve{\rho_1}, \breve{\rho_2} )$.
\end{itemize}

\subsubsection{Variance}
Even if the convergence's properties of the Maximum Pseudo-Likelihood estimator is well known under specified conditions discussed above, the variance of the estimator has to be carefully estimated. 
For the variance-covariance matrix of the coefficients, we propose to compute the matrix $\mathbf{U}'\mathbf{W}\mathbf{U}$ as if we were in the case of a classic variance in the case of maximum likelihood in the logistic case. The matrix $\mathbf{U}$ is a $nT \times p$ matrix defined by theses rows $\mathbf{U}_{it,.}=(1,X_i^T,\sum_{j \in {\cal N}_i }Y_{jt},Y_{i,t-1})$ for each $(i,t)$, $1 \leq t \leq T$ and $1 \leq i \leq n$.  $\mathbf{W}$ is the diagonal $nT \times nT$ matrix with  coefficients being equal to $\breve{p}_{it}(1-\breve{p}_{it})$ that depends on the estimation parameters $( \breve{\boldsymbol{\beta}},\breve{\rho_1}, \breve{\rho_2} )$.

In the simulations, we compared the variances of the estimators with ones computed by "bootstrap".

\subsection{Model choice}
Although the method of estimation works well for a given neighborhood structures, it is necessary to select the best one to properly fit the real dataset.

To be honest, at first instance, we had the idea of adapting the ABC method  (for Approximate Bayesian Computation) proposed by \cite{grelaud2009abc} in order to choose the best model for the neighborhood. ABC  is a likelihood-free inference method in the bayesian framework that is very convenient  when the likelihood is not available in a closed form. 
First introduced by \cite{pritchard1999population} and expanded in \cite{beaumont2002approximate} and \cite{marjoram2003markov}, ABC method was adapted by Grelaud {\it et al.} \cite{grelaud2009abc} for model choice in Gibbs random fields (GRF). We first think of this method because our model is not so far from a GRF. But because of the centering parametrization, it is not possible to exhibit a exhaustive statistics in order to compute a simple  distance between the simulated field and the observed one. We can see in Equations of the spatial joint laws of \autoref{th:exist}, that the parameters $(\boldsymbol{\beta},\rho_1)$ are embedded in the potentials depending on the spatio-temporal field $ \mathbf{Z}$. We  tried this method with different statistics without success.  However, this sophisticated method is not essential here because we observed from the simulations that the Loglikelihood is a very simple and performant indicator to choose the model of a given data set. 

The principle we propose is the following: we use "experts opinions" to determine a set of possible neighbors graphs to consider. For each graph, we perform the estimation of the parameter and simply choose the graph that optimize the log-likelihhod.

\subsection{Simulation study}

\subsubsection{Model 1}
The first configuration was again $20*20$ grid for 15 years without covariate. The initial field is generated by Bernoulli distribution with parameter 0.1 and parameters are $\beta_0=-1.4, \beta_1=0, \rho_1=0.5, \rho_2=0.5$.

\begin{figure}%
\begin{center}
\includegraphics[width=0.5\textwidth]{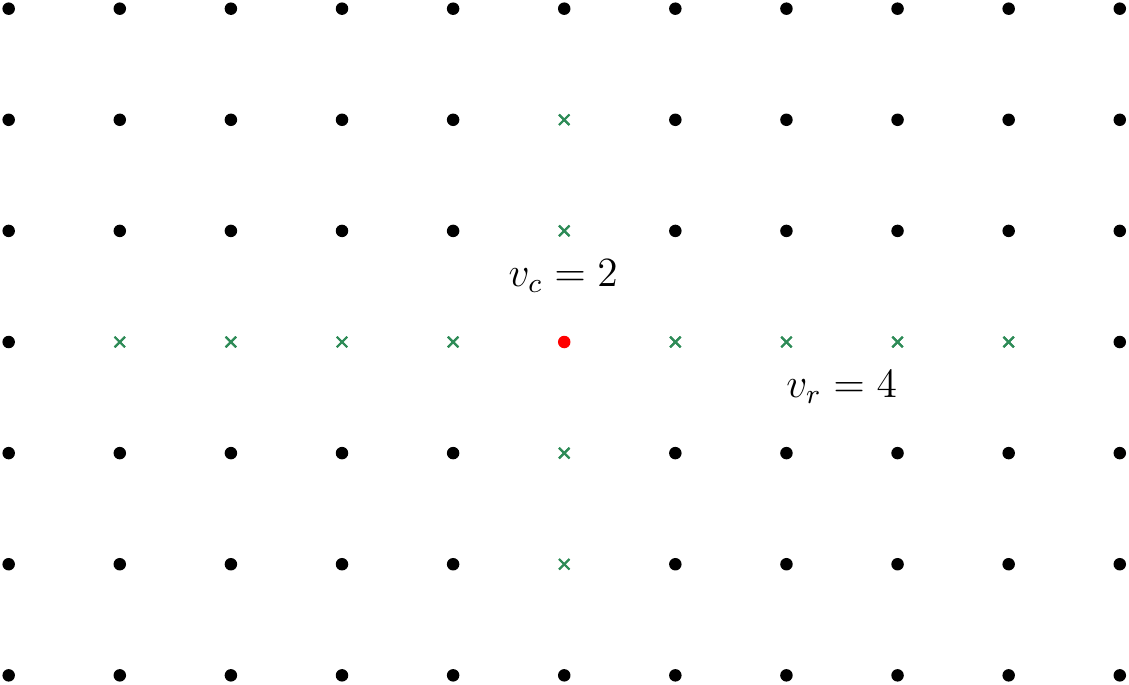}%
\end{center}
\caption{Structure of neighborhood used for the simulations. Green crosses are the neighbors of the red point.}
\label{croix}%
\end{figure}

We standardize the neighborhood structure here: suppose the points i's are on a grid, and we assume their spatial distribution is like a matrix -- each point is located on the intersection of a row and a column.   All the simulations performed in the previous section have the same structure of neighborhood: every point admits four neighbors on the same row (the four nearest that is in our case the two nearest on each side) and also two neighbors in the same column. Points of the first row (resp. second row) have only two (resp. three) neighbors on the same row and points on the first and second columns have only one neighbor on the same column (and so on for the last row or column). This configuration is denoted $v_r=2$ (for 2 neighbors on each side in the row) and $v_c=1$ (resp. 1 in the column). Figure \ref{croix} exhibit such kind of neighborhood for $v_r=4$ and $v_c=2$ 
 
  For our simulation study to infer the neighbor's struture, we make the hypothesis that all the considered structures of neighborhood are regular that is, it follows the same rule for each point of the grid. And we only consider models defined by their number of neighbors on each side in the same row  $v_r$  (resp. on the same column $v_c$). We do not consider the possibility to have neighbors on the diagonal in our simulations.

\subsubsection{Model 2 }

Again a  $20*20$ grid for 15 years with one temporal covariate with large variation first increasing from $1$ to $8$ the $8$ first years and next decreasing by $1$ until $1$ the year $15$. $x(t)=t$ for $1\leq 8$ and $x(t)=14-t$ for $1\leq 7$, $\beta_0=-2.8, \beta_1=0.1, \rho_1=0.5, \rho_2=0.5$. The structure of neighborhood is given by $v_r=2$ and $v_c=1$.

\subsubsection{Estimation results }
We used again PGS sampling methods described in section \ref{ss:SA} to perform simulations
 of a $20*20$ grid  for 15 years in different configurations to see the performance of the  estimations and the procedure to choose the model.  We compute the estimations via EMPL algorithm detailed in section \ref{sec:EMPL}. 
 
For the variance, we compared the values estimated by the method explained above on one sample with the experimental value of the variance when we estimate $B=100$ independent repetitions of the same model with the same parameters. Results of the inference for Model 1 (resp. Model 2) is given in table \ref{t:res1} (resp. \ref{t:res2}).

\begin{figure}%
\includegraphics[width=0.4\textwidth]{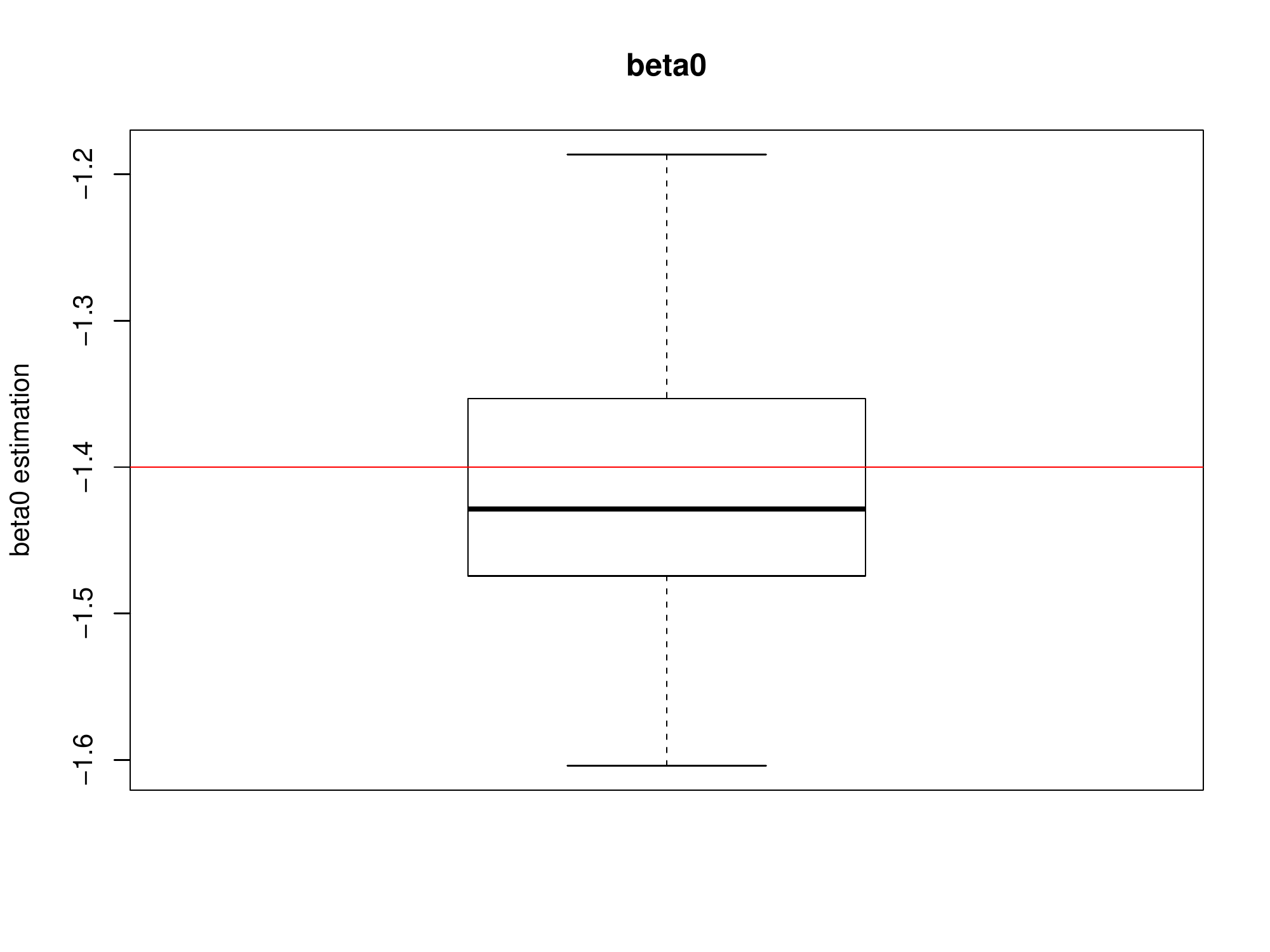}%
\includegraphics[width=0.4\textwidth]{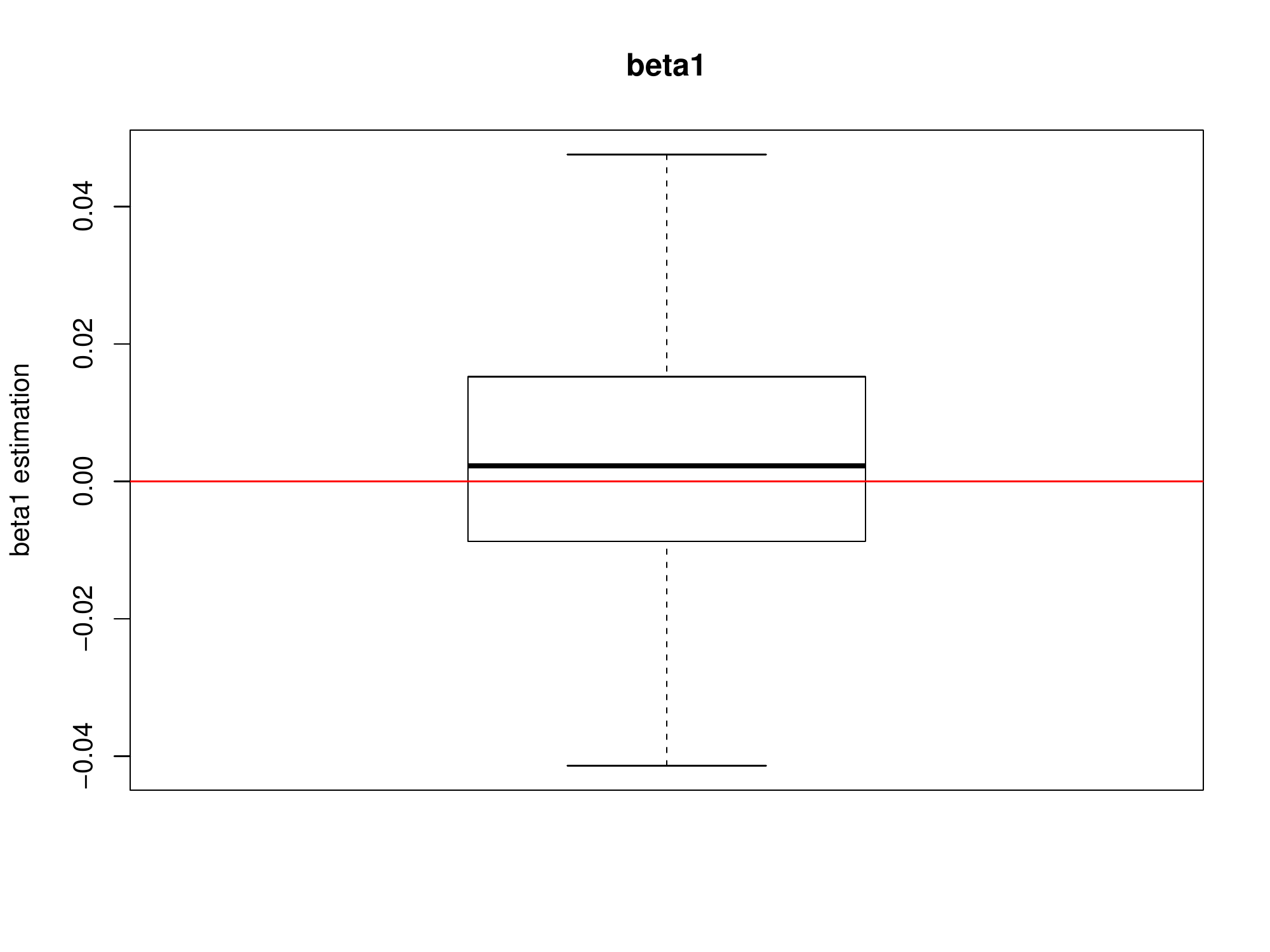}\\
\includegraphics[width=0.4\textwidth]{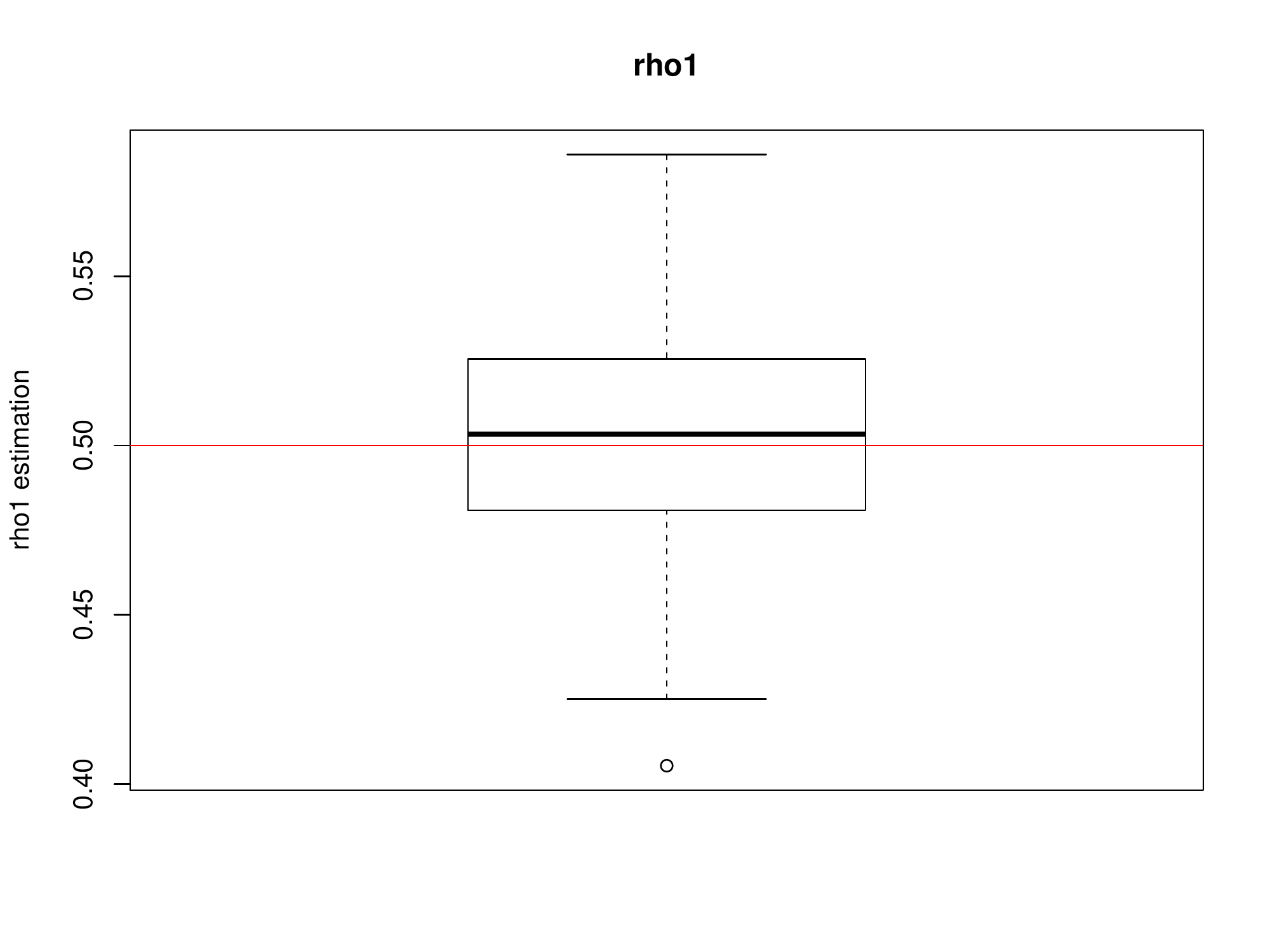}%
\includegraphics[width=0.4\textwidth]{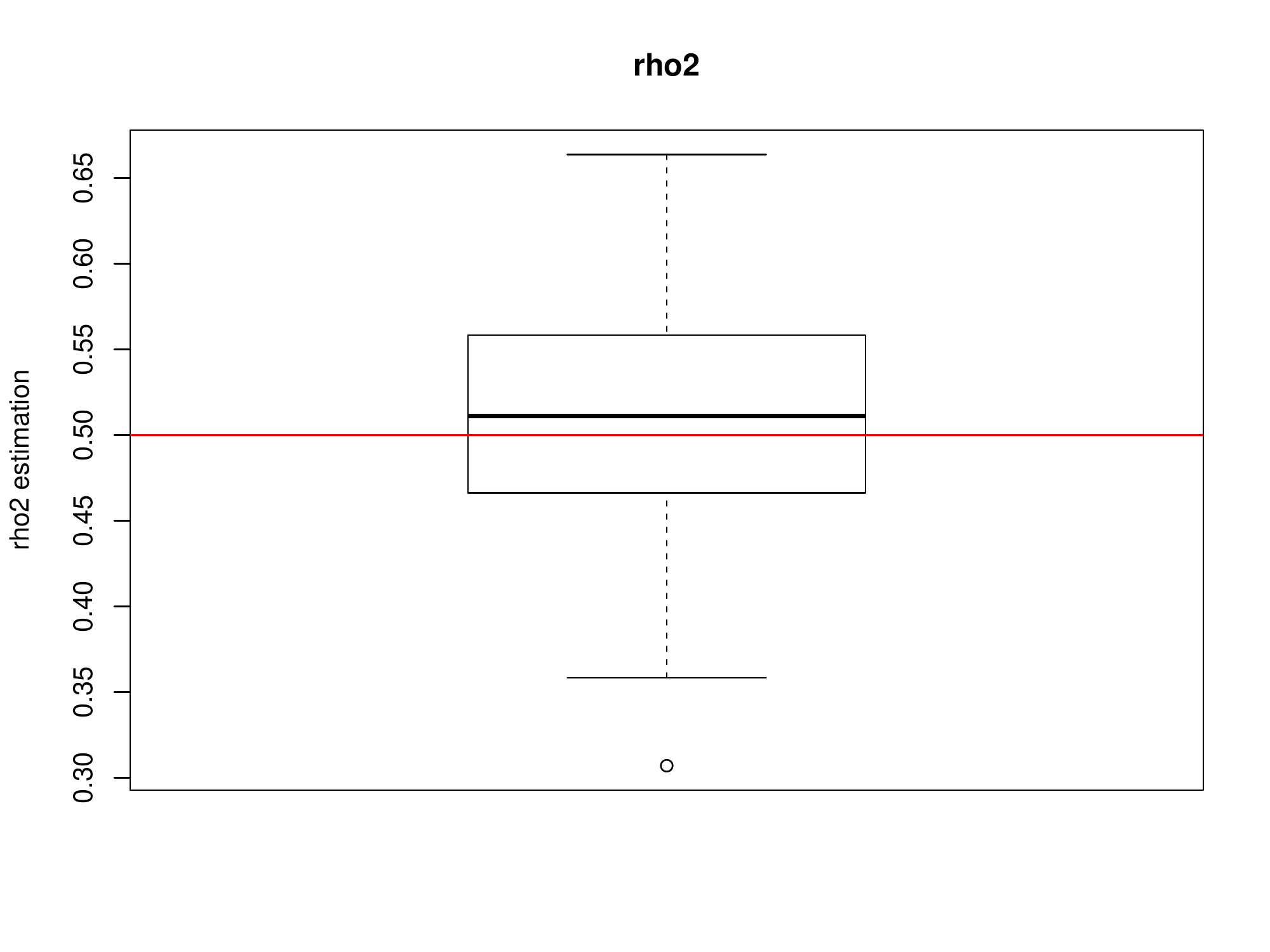}
\caption{Box-plot of $B=100$ estimations of the four parameters in Model 1. Red lines show the true values of parameters.}
\label{ML}%
\end{figure}

\begin{figure}%
\includegraphics[width=0.4\textwidth]{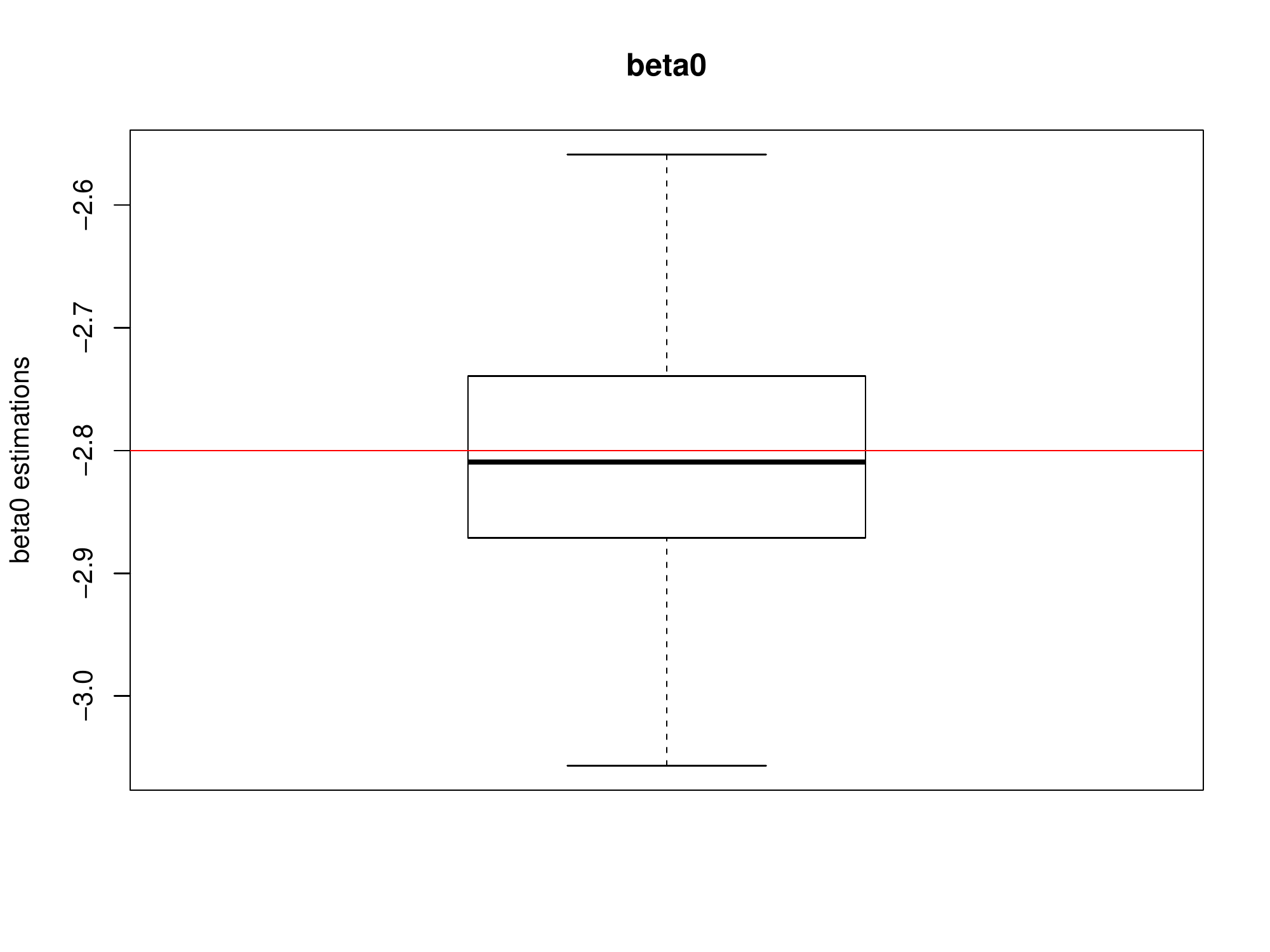}%
\includegraphics[width=0.4\textwidth]{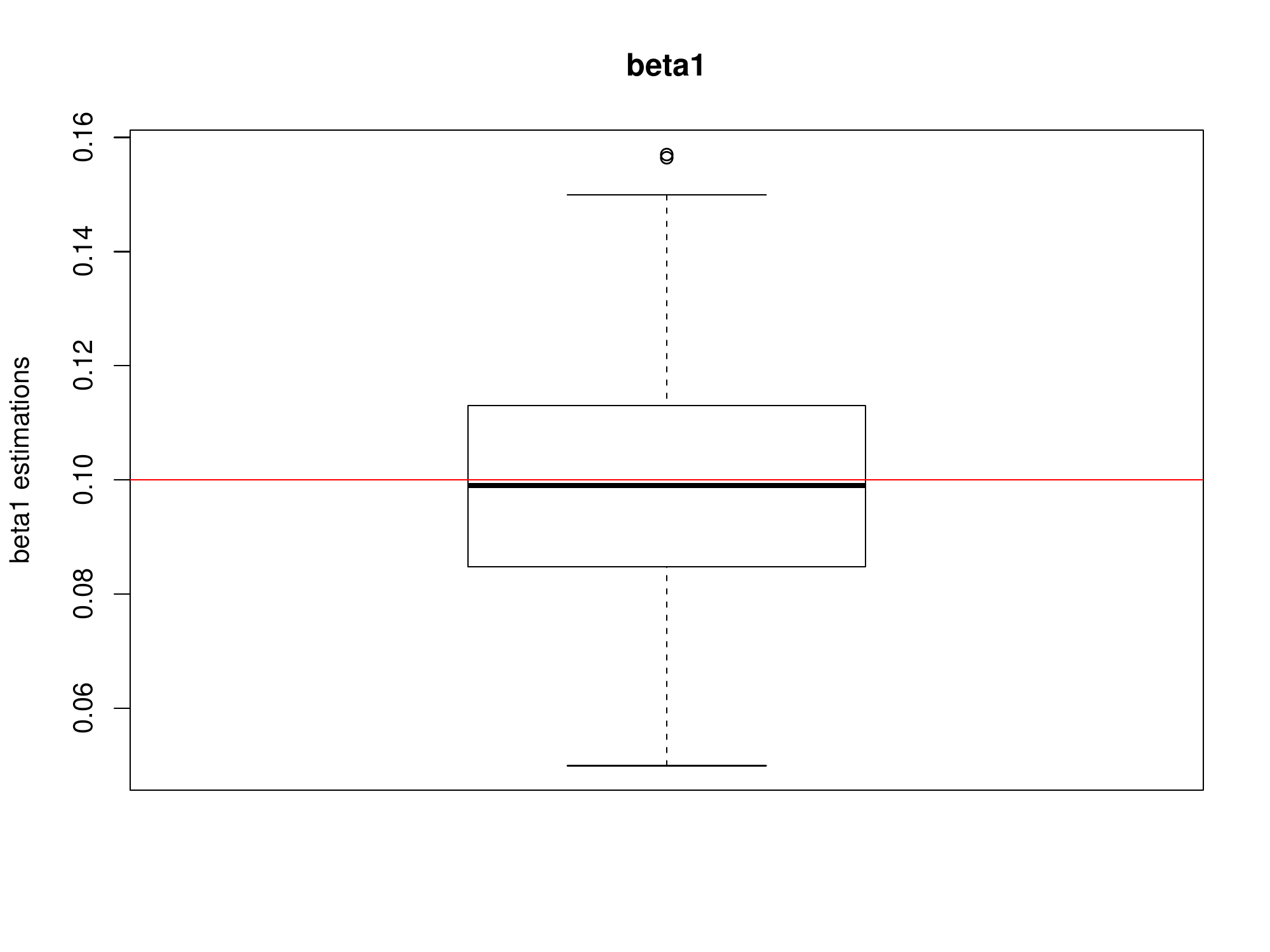}\\
\includegraphics[width=0.4\textwidth]{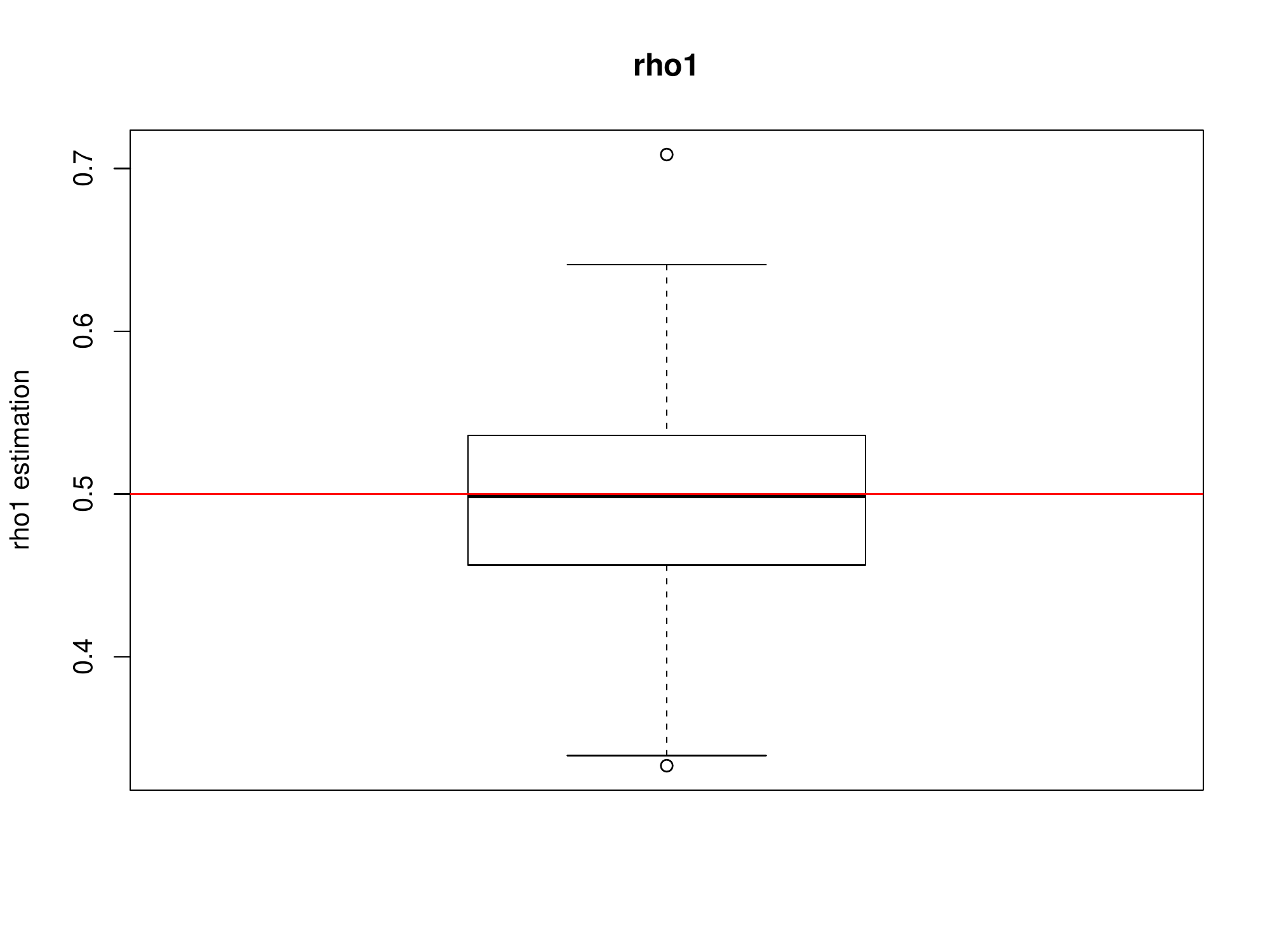}%
\includegraphics[width=0.4\textwidth]{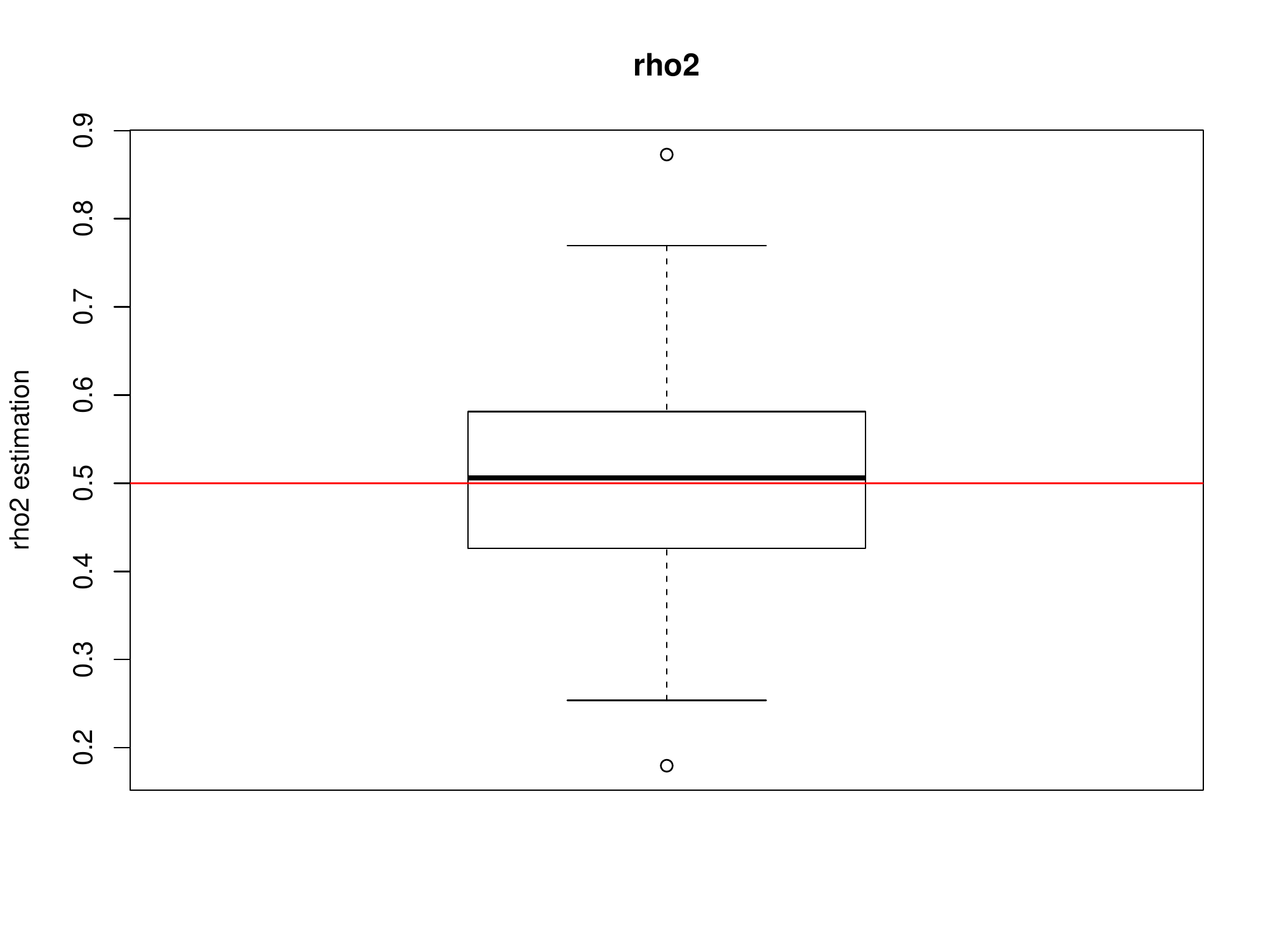}
\caption{Box-plot of $B=100$ estimations of the four parameters in Model 2. Red lines show the true values of parameters }
\label{MLx}%
\end{figure}

 Figures \ref{ML} and \ref{MLx} show the dispersion of $B=100$ estimations of $B$ independent simulations of each model. 

{\small
\begin{table}[h]
\begin{tabular}{lllll}
   \hline
   \hline
        & $\beta_0$& $\beta_1$ & $\rho_1$  & $\rho_2$ \\
mean     & -1.47(-1.4)& 0.003(0) & 0.519(0.5)  &0.560(0.5)  \\
est st.deviation & 0.066 &0.014 & 0.028 & 0.071 \\
boot. st.deviation & 0.083 &0.018 & 0.034 & 0.068 \\
   \hline
   \end{tabular}
\caption{Maximum Pseudo-Likelihood estimation for Model 1 without covariate, true values are in brackets. Variance estimators is computed by our proposed method (est st.estimation) and by repetitions method (boot st.estimation)}
\label{t:res1}
\end{table}}

{\small
\begin{table}[h]
\begin{tabular}{lllll}
   \hline
   \hline
         & $\beta_0$& $\beta_1$ & $\rho_1$  & $\rho_2$ \\
mean     & -2.757(-2.8) & 0.094(0.1) & 0.488(0.5)  &0.486(0.5)  \\
st.deviation & 0.097 & 0.021  & 0.042 & 0.130 \\
boot. st.deviation & 0.108 &0.022 & 0.073 & 0.130 \\
   \hline
\end{tabular}
\caption{Maximum Pseudo-Likelihood estimation for Model 2 with temporal covariate, true values are in brackets. Variance estimators is computed by our proposed method (est st.estimation) and by repetitions method (boot st.estimation).}
\label{t:res2}
\end{table}}

All these results show the good performance of the Maximum Pseudo-Likelihood estimator and of the standard deviation of them. We have to note that the method is very easy to implement and all those results are available almost instantaneously. It has to be noted that it would be not the case with MCMC or Bayesian methods. Note also that we performed simulations and estimations for other values of the parameters and that the performances of this method were the same.

\subsubsection{Model choice results}

We first show the effects of different choices of neighborhood graphs on the estimation of the spatial auto-regression parameter $\rho_1$ of equation (\ref{eq:model}).  For a given simulated dataset, we perform estimations of the parameters for different graphs of neighborhood. Results are shown in \autoref{voisin} that confirms the intuitive result that the estimation of the spatial autoregressive parameter decreases with the number of neighbors. We have to notice that this decrease is not proportional to the number of neighbors of each points of the grid. Moreover estimation with a wrong neighborhood structure does not affect the estimation of the other parameters of the models (regression on the past and on the covariate).

To see the good performance of the model choice rule, we simulate $100$ independent realizations of different kinds of models under three different neighborhood structures, without (resp. with a temporal covariate) and for three different values of $\rho_1$. We estimated the parameters under  six different neighbors structures and choosed the model with the biggest Pseudo-likelihood. Results are shown in \autoref{vois1} (resp. \autoref{vois2}).

{\small
\begin{table}[h]
\centering
\begin{tabular}{cll|llll}  
   \hline
   \hline
Model &  $v_r$ & $v_c$     & $\beta_0$& $\beta_1$ & $\rho_1$  & $\rho_2$ \\
   \hline
1& 1 & 1     & -1.434 & 0.005 & 0.604 &0.539  \\
 {\it 2} & {\it 2} & {\it 1}     & {\it -1.470(-1.4)}& {\it 0.003(0)} & {\it 0.519(0.5)}  &{\it 0.560(0.5) } \\
3 &2 & 2     & -1.467 & 0.003 & 0.420  &0.560  \\
4 & 3 & 1     & -1.468 & 0.004 & 0.427  &0.563  \\
5 &3 & 2     & -1.475& 0.004 & 0.368  &0.563  \\
6 &3 & 3     & -1.464& 0.004 & 0.317  &0.567  \\
   \hline
\end{tabular}
\caption{Estimation by Pseudo-Likelihood in different models for neighborhood. $v_r$ (resp. $v_l$) is the number of neighbors on each side of a point on the raw (resp. on the column).  Estimations for the true model are in red and the true values of the parameters are in brackets.}
\label{voisin}
\end{table}}


{
\begin{table}[h]

\centering
\scriptsize
\begin{tabular}{c||cccccc||cccccc||cccccc||}

   \hline
   \hline
       & \multicolumn{18}{c}{ Chosen Model} \\
   \hline 
   True   & \multicolumn{6}{c||}{$\rho_1=0.3$}& \multicolumn{6}{c||}{$\rho_1=0.4$}& \multicolumn{6}{c||}{$\rho_1=0.5$}\\
 Model   & 1    & 2& 3& 4& 5 & 6& 1    & 2& 3& 4& 5 & 6& 1    & 2& 3& 4& 5 & 6 \\
\hline
  1   & 91    & 6& 1& 2& 0 & 0& 99   & 0& 1& 0 & 0 & 0 & 100    &  0&  0& 0 & 0 & 0 \\

 2 &1& 90& 6& 2& 0 & 1& 0    & 99 & 0& 1& 0& 0& 0    & 100 &0& 0 & 0 & 0 \\

  3     &0  &3& 92& 0& 4& 1 & 0    & 0& 98& 0& 2 & 0&  0    & 0& 100& 0 & 0 & 0 \\
\hline
\end{tabular}
\caption{Model choice by maximizing the Pseudo-Likelihood for a model on a $20*20$ grid  for $15$ years without covariate. $\beta_0 = -1.4$, $\rho_2 =  0.5$ and three different values of $\rho_1$.}\label{vois1}
\end{table}}

{
\begin{table}[h]

\centering
\scriptsize
\begin{tabular}{c||cccccc||cccccc||cccccc||}

   \hline
   \hline
       & \multicolumn{18}{c}{ Chosen Model} \\
   \hline 
   True   & \multicolumn{6}{c||}{$\rho_1=0.3$}& \multicolumn{6}{c||}{$\rho_1=0.4$}& \multicolumn{6}{c||}{$\rho_1=0.5$}\\
 Model    & 1    & 2& 3& 4& 5 & 6& 1    & 2& 3& 4& 5 & 6& 1    & 2& 3& 4& 5 & 6 \\
\hline
  1    & 67   & 13& 8& 8& 2 & 2& 81   & 8& 6&3 &1 & 1 & 92    &  6&  0& 2 & 0 & 0 \\

 2 & 10    & 53& 13& 20& 1 & 3& 9    & 71 & 9& 9& 0& 2&2    & 83&8& 7 & 0 & 0 \\

  3    &4  &10& 60& 1& 14& 11 & 1    & 7& 87& 0& 3 & 2&  1    & 6& 90& 0 & 3 & 0 \\
\hline
\end{tabular}
\caption{Model choice by maximizing the Pseudo-Likelihood for a model on a $20*20$ grid  for $15$ years with a covariate. $\beta_0 = -2.8$, $\beta_1=0.1$, $\rho_2 =  0.5$ and three different values of $\rho_1$. $X_t= t$ for $t \leq 8$ and $16-t$ for $8 \leq t \leq 15$.}
\label{vois2}
\end{table}}

%
%
%
%

We see that the ability of the rule to detect the true neighbors structure is globally good. However, it has better performance if the model does  not include covariates and almost perfect while $\rho_1 \geq 0.3$. Note that if the value of $\rho_1$ is low, it means that the spatial autocorrelation with the neighbors is low and thus it is not so important to infer the neighborhood structure properly.  The performance of the rule is degraded by the presence of a covariate and again when the relative weight of the spatial auto-correlation in the model is lower. 

\section{Application to real data}
Even if the spectrum of applications of spatio-temporal autologistic models is large (see for instance the nice application for the inference of network in the purpose to infer competition in financial markets  \cite{betancourt2018investigating}) we have builded our model in the purpose to apply it in tyhe framework of vegetal epidemiology. The purpose here is to analyze the spread of an disease called esca during 14 consecutive years in a vineyard of Bordeaux in France, including 1 980 vines in a block of 30 rows and 66 columns. Esca is a grapevine trunk disease that remains poorly understood but causing extensive damage in vineyard worldwide and resulting in major economic losses (Bretsch et al. 2013; Mugnai et al. 1999)  \cite{bertsch2013grapevine, mugnai1999esca}. 
Grapevine esca is  a complex dieback disease associated with pathogenic fungi that degrade the woody part of the vine. It  exhibits discolored foliar symptoms (Mugnai et al. 1999; Surico et al. 2008) \cite{surico2008esca}. The foliar symptoms are erratic and can appear one year and disappear the next one.
 To study the spatio-temporal dynamic of the esca at the scale of a vineyard, Stefanini et al. (2000) \cite{stefanini2000longitudinal} proposed a non-centered auto-logistic multinomial statistical model with autoregression on the past and on the neighborhood.  They did not discuss nor the modelling  neither the inference method. More recently, Zanzotto et al. (2013) \cite{zanzotto2013spatiotemporal}  analyzed 17 years of data from one vineyard surveyed from planting. They use also a non-centered autologistic model with bayesian inference. In \cite{li2016spatial}, we used join count procedures to analyse aggregation and spread of the Esca 
  during a period of eight years.

\begin{figure}%
\begin{center}
\includegraphics[width=0.5\textwidth]{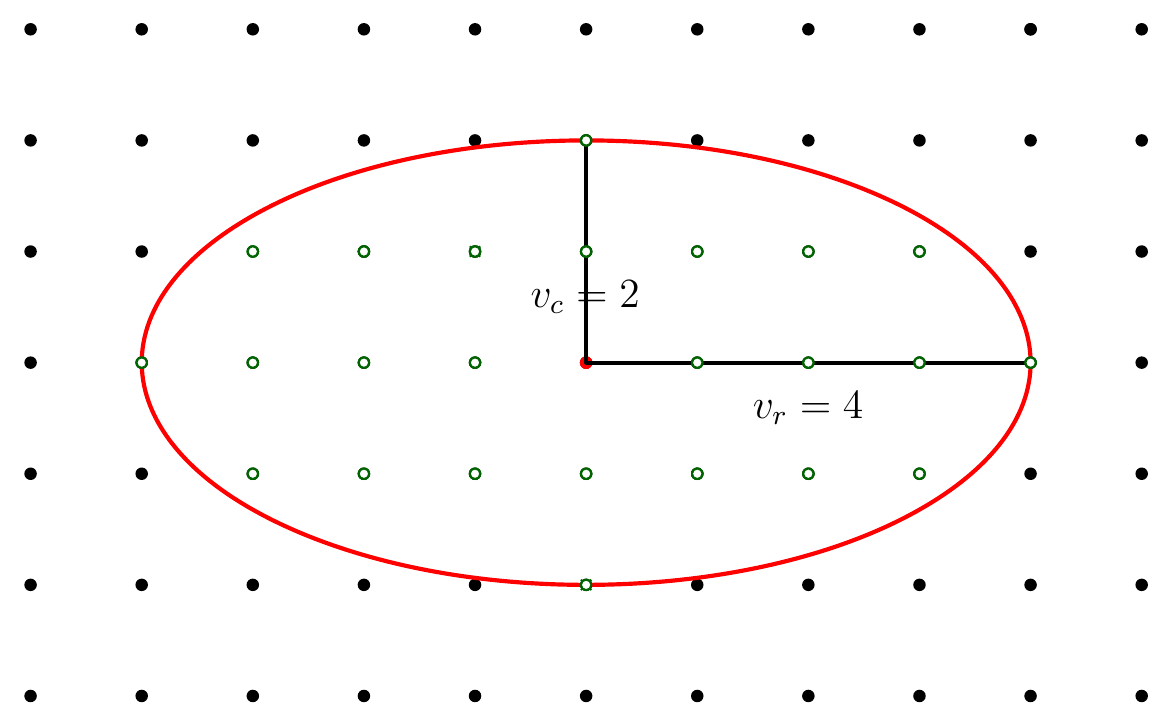}%
\end{center}
\caption{Structure of neighborhood used for esca data. Green circles are the neighbors of the red point.}
\label{ellipse}%
\end{figure}

We propose to analyse our data  in the objective to understand: (i) the effect of the status of a vine (symptomatic or not) the year preceding the observation (ii) the effect of the observed number of infected plants among the neighbors of a vine the year preceding the observation on the occurrence of the symptom for the given vine. Moreover we want to capture the instantaneous correlation between vines in the same year. We want to choose the best neighborhood structure for the effect of the previous year and the best one for the instantaneous effect. We are clearly in a context of choice of models concerning the structures of the past and instantaneous neighborhood.
After discussion with physiopathologist, the shape of neighborhood structures is given by ellipses (see \autoref{ellipse}). Indeed, vines are cultivated along the rows of the vineyard leading to possible anisotropy.   The neighborhood of a location $i$ is given by all vines included in a ellipse defined by its semi-major axis and semi-major axis. These quantities are denoted $v_r$ and $v_c$ for the instantaneous neighborhood ${\cal N}_i$ (resp.  $p_r$ and $p_c$ for the past one ${\cal N}_i^p$).The model is  the following. For a vine $i$ at time $t$, the probability to present the symptom according to the history of the vineyards and the neighborhood is given by: 

\begin{equation}
\label{eq:real}
logit(p_{it})= \beta_0 + \beta_1 \sum_{j\in N^p_i}Z_{j,t-1}+    \rho_1\sum_{j\in N_i}Z^{\ast \ast}_{j,t}+ \rho_2 Z_{i,t-1}
\end{equation}

We have tested 25 neighbor structures for the two auto-regressions (instantaneous and on the past) corresponding to $625$ different models. We define a neighborhood by an ellipse around the vine with a distance $v_r$ (resp. $v_c$) in the direction of the row (resp. column). $v_r$ and $v_c$ vary independently from $1$ to $5$ meter  leading to 25 possible structures. We  also define $p_l$ and $p_c$ the corresponding parameters for the neighborhood concerning the past. 

The most important result is that the value of the estimation of coefficient $\rho_2$ is very robust whatever neighborhood structure chosen, we found  $\breve{\rho_2} = 2.28$. It means that the probability for a vine that has already expressed the symptoms the previous year to express them again is multiplied by $ \exp(2.28) = 9.7$ comparing to another without expression the previous year. 

About the other terms of the regression, another important result is that the choice of the neighborhood for the instantaneous correlation  is more important  than the one for the past regression. The 50 best models according to the PL are the ones with the instantaneous neighborhood define by an ellipse (that also could be a circle) with radius of 5 in a direction of the row and 4 on the other. This effect is more important than the effect on the past that cover all the possibilities for the neighborhood. The interpretation of this instantaneous autocorrelation is probably that it is sourced by a local effect of environment like the properties of the soil. This effect is not due to a spread of the illness. Note that it captures a little anisotropy showing more effect along the row, that is a standard result in vineyards because the vines were nursed along the row.  
We give in \autoref{real} the estimated coefficients with their standard deviation for the best model. 

{\small
\begin{table}[h]

\centering  
\begin{tabular}{ ll|ll|llll}  
   \hline
   \hline
$v_r$ & $v_c$     &   $p_r$ & $p_c$     & $\beta_0$& $\beta_1$ & $\rho_1$  & $\rho_2$ \\
   \hline
 5 & 4     & 1&1 & -3.04 (0.035) & 0.178  (0.034) & 0.135 (0.006) &2.28 (0.05)  \\
   \hline
\end{tabular}
\caption{Instantaneous and past neighbor structures and estimated coefficients for the best model on the real data. Standard deviation of the parameters are in bracket.}
\label{real}
\end{table}}
We have already commented the estimation of autoregression $\rho_2$. The value of the one for $\beta_0$ indicates a spontaneous level of infection equal to $\exp(-3,04) =0.05$. $\beta_1$ is the coefficient of regression that quantifies the spread of the illness, like in the logistic model, when the level of the illness is still low, the presence of a one more vine with symptoms in the neigborhood at a time multiply the risk to present symptoms by $\exp(0.178)= 1.19$ the following year.   $\rho_1$ measures the spatial autocorrelation but its value is not easily interpretable. However, it shows a positive autocorrelation between occurrence of illness in the same neighborhood.

\section{Discussion and conclusion}
In this paper, we have proposed a new spatio-temporal model for the following of binary data on a lattice. At each time, the spatial covariates are centered by its expected value that depends on the value of the covariates and also on the values of the field in the past. Simulations study shows the interest of the centering particularly for the interpretation of the spatial regression parameter. We have shown the ability of the maximum pseudo-likelihood estimator to infer quickly the value of the parameters. Maximizing pseudo-likelihood allows also to choose efficiently between different structures of neighborhood. Even if the law of the whole spatio-temporal joint law of the process is not proved to exist, we still discuss the existence of the spatial joint law of the process at each time given the covariates and the past of the process. This point of view seems coherent with the recursive construction of the process along the time.

Even if we propose a method to estimate the variance of the estimators, this last point has probably to be improved.

Whatever, the model and the method proposed in this paper, are very suitable and efficient to model the evolution of an illness on a lattice taking into account for covariates and for the spatial auto-correlation. It allows to measure and quantify effects of the neighborhood in the past on the occurrence of the illness at a given time. 

\section*{Acknowledgements}
We wish to thank Avner Bar Hen and Cécile Hardouin for precious discussions at the beginning of this work. 
We wish to acknowledge the vine-grower who participated in this study and also  Sylvie Bastien and David Morais for their excellent technical assistance. This reserach was supported  by Bordeaux Sciences Agro, the Regional Council of Aquitaine, the JEAN POUPELAIN Foundation, the French Ministry of Agriculture and the Food-processing industry and Forest (CASDAR V1303).






\section*{References}
\bibliographystyle{plain} 
\bibliography{biblio-thesis}


\end{document}